\documentclass[%
 reprint,
 amsmath,amssymb,
 aps,
]{revtex4-2}

\usepackage{graphicx}
\usepackage{dcolumn}
\usepackage{bm}
\usepackage{xcolor}
\usepackage{siunitx}
\usepackage{ragged2e}
\makeatletter
\renewcommand\@make@capt@title[2]{%
  \@ifx@empty\float@link{\@firstofone}{\expandafter\href\expandafter{\float@link}}%
  {\textbf{#1}}\@caption@fignum@sep\justifying #2\quad
}%
\makeatother
\usepackage{physics}
\usepackage{enumitem}

\begin{document}

\preprint{APS/123-QED}

\title{High-fidelity conditional-squeezing gate in superconducting circuits via refocusing}

\author{Roman Schiaffino}
 \email{r.schiaffino@df.uba.ar}
\author{Fernando C. Lombardo}%
 \email{lombardo@df.uba.ar}
\author{Juan Pablo Paz}%
 \email{paz@df.uba.ar}
\affiliation{%
 Departamento de F\'isica, Facultad de Ciencias Exactas y Naturales, Universidad de Buenos Aires, Buenos Aires, Argentina
}%
\affiliation{%
 Instituto de F\'isica de Buenos Aires (IFIBA), CONICET -- Universidad de Buenos Aires, Buenos Aires, Argentina
}%

\begin{abstract}
We propose a conditional-squeezing gate for superconducting circuits, which squeezes one or two modes of a resonator along a direction determined by the state of a dispersively coupled qubit. We show that the fidelity of this gate can be substantially improved by a refocusing protocol, in an architecture accessible with current technology whose key element is a SQUID-terminated resonator. The refocusing protocol relies on a symmetry of the Hamiltonian that is absent in the previously proposed controlled-squeezing gate, which squeezes the resonator only for one state of the control qubit and, ideally, acts trivially otherwise. We apply the refocused gate to encode an arbitrary qubit state into superpositions of single- and two-mode squeezed vacuum states, in which single-photon losses are detectable through parity measurements. Including nonlinearities and dissipation, the conditional gate consistently outperforms the controlled-squeezing gate, and, for optimistic but achievable experimental parameters, the encoding reaches fidelities above 0.99 for all qubit states.

\end{abstract}

\maketitle

\section{Introduction}
The ability to control and manipulate quantum systems for quantum information processing has made unprecedented strides in the past decade. Conventional schemes that rely on information encoding in ensembles of two-dimensional (qubit) systems continue to serve as a testbed for quantum error correction \cite{Google2023, Google2024, Bluvstein2024}, and fault-tolerant, universal quantum computation \cite{Zhang2025, Butt2026} and simulation \cite{Lamata2026, Kim2023}. An alternative approach, which circumvents the need for multiple physical qubits, is to redundantly encode information in the infinite-dimensional Hilbert space of a bosonic mode, also referred to as a continuous-variable (CV) system. The seminal work of Lloyd and Braunstein \cite{PhysRevLett.82.1784} laid the theoretical foundations for using CV systems to perform universal quantum computation, a proposal followed by extensive experimental efforts toward its implementation \cite{ Pittman2005,  Madsen2022, Leindecker2026, Lienhard2025}.

Within this landscape, circuit quantum electrodynamics (cQED)~\cite{Blais2021} has emerged as a leading platform for CV quantum information processing~\cite{Joshi2021}. Its main strength is the readily available nonlinearity provided by Josephson junctions, non-dissipative circuit elements whose nonlinear inductance can be engineered by circuit design. Josephson junctions are the building blocks of artificial atoms, such as transmon qubits, that interact with microwave photons stored in long-lived superconducting resonators. They can also be embedded in the resonators themselves, as in SQUID-terminated resonators. These tunable nonlinear circuits have enabled the observation of fundamental quantum phenomena, such as the dynamical Casimir effect~\cite{Wilson2011} and multipartite non-Gaussian entanglement~\cite{4ts4-qj74}. Their nonlinearity is also a key resource for universal control of bosonic modes, since Gaussian operations alone are not sufficient~\cite{PhysRevLett.82.1784}.

Several techniques for universal control have been developed in cQED, depending on how the nonlinearity is introduced. Nonlinearities intrinsic to the bosonic mode, as in resonators terminated by a SQUID or a SNAIL~\cite{Sivak2019, Hillmann2020}, have been exploited to stabilize and operate Kerr-cat qubits~\cite{Grimm2020} and, more recently, to achieve universal control of a bosonic mode through drive-activated cubic interactions~\cite{Eriksson2024}. Alternatively, the nonlinearity can be provided by the dispersive interaction with an ancillary transmon. This is the mechanism underlying the selective number-dependent arbitrary phase (SNAP) gate~\cite{Heeres2015}, which has been used, for instance, to generate cubic-phase states~\cite{Kudra2022}, and it has also enabled the implementation of a universal gate set on a logical qubit encoded in superpositions of coherent states using optimal control pulses~\cite{Heeres2017}. Another key example is the conditional-displacement (CD) gate, which displaces the oscillator in opposite directions depending on the qubit state~\cite{Eickbusch2022}. Together with single-qubit rotations, the CD gate enables universal control of the oscillator.

In a similar spirit to the CD gate, Ref.~\cite{PhysRevA.111.042606} proposed a controlled-squeezing gate, which applies a squeezing operation to the resonator only if the qubit is in the state $\ket{1}$ and acts trivially if the qubit is in the state $\ket{0}$. Ref.~\cite{PhysRevA.111.042606} showed that, combined with single-qubit rotations and Gaussian operations on the resonator, this gate provides universal control.

In this paper, we propose a conditional-squeezing gate, which squeezes the resonator for both qubit states, along a phase-space direction determined by the state of the qubit. This operation is closely related to the controlled-squeezing gate and is also universal, since conjugating a displacement with it yields a conditional displacement of tunable amplitude. Having more than one universal gate that generates non-Gaussian states is useful, because each of them is better suited to different tasks. Conditional displacements provide the most natural route to, for example, cat states, whereas conditional squeezing is more natural for preparing superpositions of squeezed states, known as squeezed multiplets~\cite{4d9d-kthj}.

The implementation we discuss here considers the same superconducting circuit system used in Ref.~\cite{PhysRevA.111.042606} to implement the controlled-squeezing gate, namely a SQUID-terminated resonator dispersively coupled to a transmon qubit. In this system, both gates are affected by time-dependent terms of the same origin in the Hamiltonian, which cause a deviation from the ideal operation. However, as we will show, the conditional-squeezing gate allows these spurious effects to be compensated within the gate itself by means of a refocusing protocol, a technique with a long-standing history in nuclear magnetic resonance~\cite{Hahn1950} and dynamical decoupling~\cite{PhysRevLett.82.2417}. This compensation relies on an exact symmetry of the Hamiltonian that is absent in the controlled-squeezing gate, and it constitutes the main advantage of the conditional-squeezing gate over the latter.

As a benchmark, we apply the conditional-squeezing gate to an encoding algorithm that maps an arbitrary qubit state onto logical states that are superpositions of squeezed states. Under unitary evolution, the encoding fidelity obtained with the refocused gate is practically indistinguishable from that of the ideal gate, so that the performance of the algorithm is limited by the finite squeezing of the logical states, photon losses, and nonlinearities, rather than by an imperfect compensation of the spurious terms.

This article is organized as follows. In Sec.~\ref{implementation}, we present the implementation of the conditional-squeezing gate in a superconducting circuit, derive its Hamiltonian, and study how its slowly oscillating terms produce deviations from an ideal conditional-squeezing gate. Section~\ref{refocusing_section} is devoted to the compensation of these undesired effects by means of a refocusing protocol. In Sec.~\ref{encoding_section}, we test the performance of the compensated gate by using it to encode an arbitrary qubit state onto resonator states in an error-detectable way, including dissipation and nonlinearities, and compare it with the controlled-squeezing gate of Ref.~\cite{PhysRevA.111.042606}. In Sec.~\ref{tms_section}, we extend our proposal to a conditional-squeezing gate acting on two resonator modes. Finally, in Sec.~\ref{conclusion} we summarize our results. Appendices \ref{apB} and \ref{apA} contain the derivation of the single- and two-mode Hamiltonians of the circuit implementation and a discussion of the approximations involved, and Appendix \ref{app:blockME} the method used to solve the master equation for the open system analysis.

\begin{figure}[h!]
\includegraphics[width=0.5\textwidth]{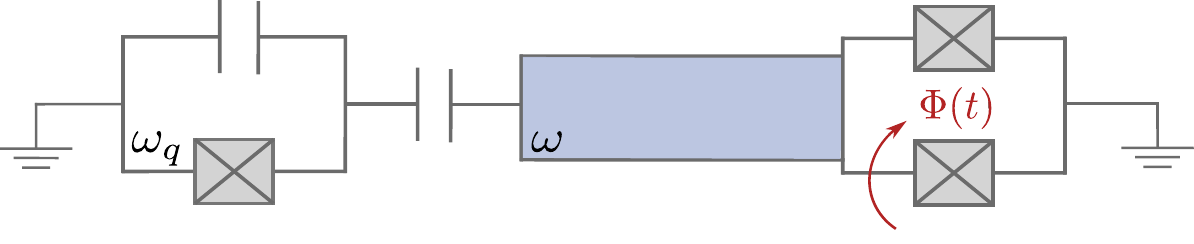}
\caption{\label{circuit}Circuit model for the implementation of the conditional-squeezing gate. A $\lambda/4$ transmission line resonator (blue) storing a single mode of bare frequency $\omega$ is terminated at one end by a SQUID, which parametrically drives the resonator through a time-dependent external flux $\Phi(t)$. The other end is capacitively coupled to a transmon qubit of frequency $\omega_q$, whose state sets the effective resonance frequency of the resonator through the dispersive interaction.}
\end{figure}

\section{The conditional-squeezing gate}
\label{implementation}

To implement the conditional-squeezing gate, we use the same setup as in
Ref.~\cite{PhysRevA.111.042606}, shown in Figure~\ref{circuit}, which comprises
three basic elements: a $\lambda/4$ transmission line resonator,
a flux-tunable SQUID terminating it, and a single-junction transmon. The transmon, which we
assume to behave as a two-level (qubit) system with states $|0\rangle$ and
$|1\rangle$ and frequency $\omega_q$, is capacitively coupled to the resonator in the dispersive regime, with dispersive shift $\chi$ (the validity of these approximations is discussed in Appendix A).
The SQUID is threaded by an external magnetic flux $\Phi(t)=\Phi_s+\delta\Phi(t)$, consisting of a static bias $\Phi_s$ and a weak time-dependent drive, $|\delta\Phi(t)|\ll\Phi_s$. We focus on a single mode of the resonator, with frequency $\omega$, and assume the other modes to be far detuned from the drive and not excited (the two-mode case is considered in Sec.~\ref{tms_section}). The static flux bias is chosen away from the zero-flux sweet spot ($\Phi_s\neq0$), so that $\bar g_d=\partial\omega/\partial\Phi|_{\Phi_s}\neq0$. The drive then modulates the resonator frequency through the inductive energy of the mode, giving rise to a parametric coupling between the driving field and the resonator which, to first order in $\delta\Phi$, is proportional to $\bar g_d\,\delta\Phi(t)$ and to the square of the field quadrature, $(\hat a+\hat a^{\dagger})^2$. Here $\hat a$ ($\hat a^{\dagger}$) is the annihilation (creation) operator of the resonator mode. Within these approximations, the Hamiltonian of the circuit reads

\begin{equation}
\label{full_hamiltonian}
\hat H(t)=\frac{\omega_q}{2}\hat\sigma_z+\omega\,\hat a^{\dagger}\hat a
+\frac{\chi}{2}\,\hat\sigma_z\hat a^{\dagger}\hat a
+\frac{\bar g_d}{2}\,\delta\Phi(t)\,(\hat a+\hat a^{\dagger})^{2},
\end{equation}

\noindent where $\hat\sigma_z=|0\rangle\langle0|-|1\rangle\langle1|$ is the Pauli operator of the qubit. A first-principles derivation of Eq.~\eqref{full_hamiltonian} from the circuit Lagrangian is given in Appendix~\ref{apB}. The Hamiltonian $\hat H(t)$ does not include nonlinearities induced by the SQUID, but we will analyze their impact in Sec.~\ref{open_system}.




The interpretation of $\hat{H}(t)$ is simple: it describes a harmonic oscillator whose frequency depends both on the qubit state (through the dispersive interaction) and on time (through the coupling with the external drive). In fact, when the qubit state is $|0\rangle$, the frequency of the resonator is given by $\omega_{0}(t)=\bar\omega_{0}+\bar{g}_d\,\delta\Phi(t)$, where $\bar{\omega}_{0}=\omega+\frac{\chi}{2}$. Similarly, when the qubit is in the state $|1\rangle$, the frequency becomes $\omega_{1}(t)=\bar\omega_{1}+\bar{g}_d\,\delta\Phi(t)$, where $\bar{\omega}_{1}=\omega-\frac{\chi}{2}$.


We consider a bichromatic drive of the form $\delta \Phi(t)=\epsilon_0\sin{(\omega_{d,0}t-\varphi_0)}+ \epsilon_1\sin{(\omega_{d,1}t-\varphi_1)}$, where $\epsilon_{j}$, $\omega_{d,j}$ and $\varphi_{j}$ ($j=0,1$) are the driving flux amplitudes, frequencies and phases, respectively. Hereafter, we assume equal driving flux amplitudes ($\epsilon_0=\epsilon_1=\epsilon$). It is simple to see that, when choosing $\omega_{d,0}=2\bar\omega_0$ and $\omega_{d,1}=2\bar\omega_1$, state-dependent parametric resonance is excited in the resonator, which is the key ingredient of the selective squeezing gate proposed here. To see this, we write the Hamiltonian $\hat H(t)$ of Eq. \eqref{full_hamiltonian} in the interaction picture with respect to its static terms (i.e., we define the free Hamiltonian as $\hat H_0=\frac{\omega_q}{2}\hat{\sigma}_z+\omega \hat a^{\dagger}\hat a+ \frac{\chi}{2}\hat{\sigma}_z\hat a^{\dagger}\hat a$). The bosonic operators oscillate in this picture at frequency $\bar\omega_0$ ($\bar\omega_1$) when the qubit is in $\ket{0}$ ($\ket{1}$), and the Hamiltonian takes the form (we define $g_d=\bar{g}_d\epsilon$)

\begin{align}
\hat{H}_I(t) &= \frac{g_d}{2}\,[\sin(2\bar{\omega}_0t-\varphi_0)+ \sin(2\bar{\omega}_1t-\varphi_1)] \nonumber\\
&\times\Bigl\{|0\rangle\langle 0|\otimes(\hat{a} e^{-i\bar{\omega}_0t}+ \hat{a}^{\dagger}e^{i\bar{\omega}_0t})^{2} \nonumber\\
&\phantom{{}\times\Bigl\{}+ |1\rangle\langle 1|\otimes(\hat{a} e^{-i\bar{\omega}_1t}+ \hat{a}^{\dagger}e^{i\bar{\omega}_1t})^{2}\Bigr\}.
\label{interaction_picture}
\end{align}

The terms proportional to $|0\rangle\langle 0|$ and $|1\rangle\langle 1|$ in Eq.~\eqref{interaction_picture} define two separate branches of the evolution. In each of them, we can write the trigonometric functions of the drive in terms of exponentials and expand the binomial term involving the resonator operators, obtaining sixteen terms. Two of them are time-independent due to the parametric resonance condition. Two other terms in each branch oscillate at frequencies $\pm 2(\bar\omega_0-\bar\omega_1)=\pm 2\chi$. The remaining twelve terms oscillate at much higher frequencies, which are of the form $\pm2(\bar\omega_0+\bar\omega_1)$, $\pm 4\bar\omega_0$, $\pm 4\bar\omega_1$, $\pm 2\bar\omega_0$ and $\pm2 \bar\omega_1$. For typical circuit QED parameters, $\chi/2\pi \sim $ $\mathrm{MHz}$ and $\omega/2\pi \sim $ $\mathrm{GHz}$, all these last twelve terms can be neglected within the rotating-wave approximation (RWA). In this case, retaining only the slowly varying terms, the Hamiltonian of Eq. \eqref{interaction_picture} reduces to

\begin{widetext}
\begin{align}
\hat{H}_I(t) &\approx |0\rangle\langle 0| \otimes \frac{(-i g_d)}{4}
\left\{
\hat a^2 e^{-i\varphi_0} - \hat a^{\dagger 2} e^{i\varphi_0}
+ \hat a^2 e^{-i(2\chi t+\varphi_1)} - \hat a^{\dagger 2} e^{i(2\chi t+\varphi_1)}
\right\} \nonumber \\
&\quad + |1\rangle\langle 1| \otimes \frac{(-ig_d)}{4}
\left\{
\hat a^2 e^{-i\varphi_1} - \hat a^{\dagger 2} e^{i\varphi_1}
+ \hat a^2 e^{-i(-2\chi t+\varphi_0)} - \hat a^{\dagger 2} e^{i(-2\chi t+\varphi_0)}
\right\}.
\label{rwa_hamiltonian}
\end{align}
\end{widetext}

Let us write the Hamiltonian in Eq. \eqref{rwa_hamiltonian} as $\hat H_I(t)=|0\rangle\langle 0| \otimes \hat H_{I,0}(t)+|1\rangle\langle 1| \otimes \hat H_{I,1}(t)$, where $\hat H_{I,j}(t)$ is the Hamiltonian acting on each branch,

\begin{align}
\hat{H}_{I,j}(t) &= \frac{-ig_d}{4}\bigl\{
\hat{a}^2 e^{-i\varphi_j} - \hat{a}^{\dagger 2}e^{i\varphi_j} \\  \nonumber 
& \phantom{{}= \frac{-ig_d}{4}\bigl\{} + \hat{a}^2 e^{-i(\Omega_j t+\varphi_{1-j})} - \hat{a}^{\dagger 2}e^{i(\Omega_j t+\varphi_{1-j})}\bigr\},
\end{align}

\noindent with $\Omega_j=(-1)^j2\chi$. Evidently, each branch Hamiltonian contains static and slowly oscillating contributions. Retaining only the static terms, the evolution operator reduces to an ideal conditional-squeezing operator, which we denote as $\text{CSq}(r,\varphi_0,\varphi_1)$ and can be written as

\begin{equation}
\label{ideal_csq}
    \text{CSq}(r,\varphi_0,\varphi_1)=|0\rangle\langle 0| \otimes \hat S(r,\varphi_0)+ |1\rangle\langle 1| \otimes \hat S(r,\varphi_1).
\end{equation}

Here, $\hat S(r,\varphi_j)=\exp[\frac{r}{2}(\hat a^2 e^{-i\varphi_j} - \hat a^{\dagger 2} e^{i\varphi_j})]$ is the single-mode squeezing operator, with squeezing parameter of magnitude $|r|=g_dt/2$, which grows linearly in time, and squeezing direction in phase space $\varphi_j/2$. The ideal gate in Eq.~\eqref{ideal_csq} applies the squeezing operation in the resonator in a state-dependent manner, with direction set by $\varphi_0$ ($\varphi_1$) when the qubit state is $\ket{0}$ ($\ket{1}$).

\begin{figure*}[!ht]
    \centering
    \includegraphics[width=1\textwidth]{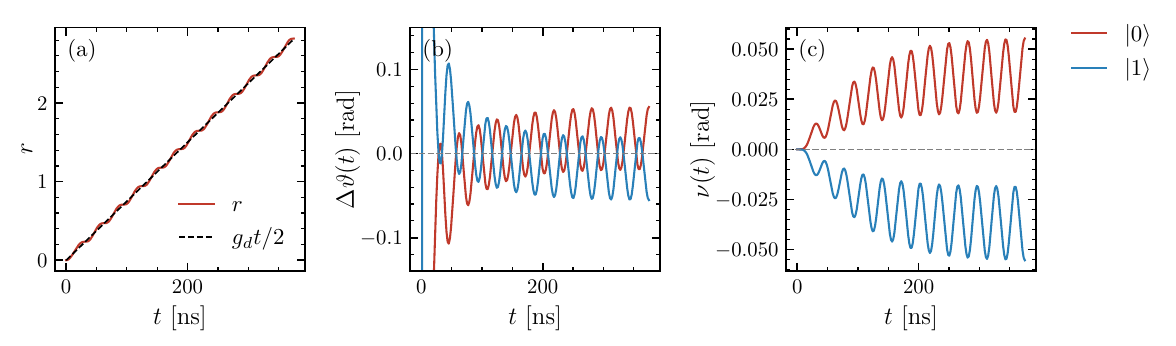}
    \caption{\label{parameters_evolution}
    Evolution of (a) the squeezing parameter $r$, (b) the quadrature deviation $\Delta \vartheta(t)$, and (c) the induced phase $\nu(t)$ of the squeezed states under the Hamiltonian $\hat H_I(t)$ up to 375 ns. For the last two, we show the evolution for the qubit states $\ket{0}$ (red) and $\ket{1}$ (blue). While the squeezing parameter does not suffer a significant deviation from the one obtained with an ideal gate (dashed line), the impact of the time-dependent terms on the squeezing direction and the induced phase is highly sensitive.}
\end{figure*}


The slowly oscillating terms retained within the RWA will clearly introduce deviations from the ideal $\text{CSq}(r,\varphi_0,\varphi_1)$ gate. Before presenting numerical results for the solution of the Schrödinger equation, we note that, since $\hat H_I(t)$ is quadratic in the resonator mode operators, the Gaussian character of the states is preserved at all times. With this in mind, it is instructive to write an initial state, such as $|\Psi_{\text{QR}}\rangle=\frac{1}{\sqrt{2}}(|0\rangle +|1\rangle)\otimes |0\rangle$, evolved under $\hat H_I(t)$ as 

\begin{equation}
\label{state_evolved}
    |\Psi_{\text{QR}}\rangle=\frac{1}{\sqrt{2}}|0\rangle\otimes|\xi_0(t)\rangle e^{i\nu(t)}+ \frac{1}{\sqrt{2}}|1\rangle\otimes|\xi_1(t)\rangle e^{-i\nu(t)}.
\end{equation}

Here, $|\xi_0(t)\rangle$ and $|\xi_1(t)\rangle$ are squeezed vacuum states whose squeezing parameter and phase-space orientation depend on time, and $\nu(t)$ is a time-dependent phase induced in the evolution. We analyze how these three parameters (namely, the squeezing parameter, the squeezing direction and the induced phase) are affected by the time-dependent terms in $\hat H_I(t)$ relative to the ideal gate.



For the simulations in this work, we use the following parameters: qubit and resonator frequencies $\omega_q/2\pi = 4$~GHz and $\omega/2\pi = 6$~GHz, a drive strength $g_d/2\pi = 2.39$~MHz, and a dispersive shift $\chi/2\pi = 16$~MHz, as in Ref.~\cite{PhysRevA.111.042606}. The drive strength corresponds to a squeezing rate $g_d/2=2\pi\times1.19$~MHz, well below the parametric beamsplitter rates of more than $2\pi\times5$~MHz demonstrated with a flux-driven DC-SQUID coupler~\cite{Lu2023_2}. 
This value of $\chi$ lies at the upper end of those typically used in cQED. We employ it here to compare with the results of Ref.~\cite{PhysRevA.111.042606}, where this value was chosen. In Sec.~\ref{encoding_section} we also consider dispersive shifts down to $\chi/2\pi=3$~MHz, scaling the drive strength accordingly. We solved the Schrödinger equation using QuTiP with a resonator Hilbert space of dimension $N=1000$. We set the drive phases to $\varphi_0=0$ and $\varphi_1=\pi$, which define the target squeezing directions (i.e., those obtained with the ideal gate): $0$ for the qubit state $\ket{0}$ and $\pi/2$ for the qubit state $\ket{1}$.

We found that the time-dependent terms in $\hat H_I(t)$ do not significantly affect the squeezing parameter of the states, which still grows linearly in time up to a small harmonic modulation of amplitude $\sim10^{-2}$, as shown in Figure~\ref{parameters_evolution}(a) for evolution times up to 375 ns. Thus, $r$ remains close to its ideal magnitude $g_dt/2$ (shown in dashed lines), and the small deviation has no practical effect on the gate performance. To characterize the squeezing direction, we introduce the quadrature deviation $\Delta \vartheta_j(t)$, defined as the difference between the instantaneous squeezing angle $\vartheta_j(t)$ and the target direction: $\Delta \vartheta_0(t) = \vartheta_0(t)$ and $\Delta \vartheta_1 (t)= \vartheta_1(t) - \pi/2$. As shown in Figure~\ref{parameters_evolution}(b), the squeezing orientation oscillates in phase space around a certain direction close to the target. The evolution of the induced phase $\nu(t)$ can be seen in Figure~\ref{parameters_evolution}(c). This phase exhibits a harmonic modulation around a linearly growing value before settling into oscillations of constant amplitude.




Both the existence of the phase $\nu(t)$ and the quadrature oscillation, caused by the time-dependent terms of $\hat{H}_I(t)$, clearly affect the conditional-squeezing gate fidelity and therefore must be compensated. In the following section, we present a method that achieves this, based on a special symmetry exhibited by $\hat{H}_I(t)$: making explicit the dependence on both drive phases, one can verify that $\hat{H}_{I,0}(t,\varphi_0,\varphi_1)=\hat{H}_{I,1}(-t,\varphi_1,\varphi_0)$. That is, $\hat{H}_{I,0}(t)$ is obtained from $\hat{H}_{I,1}(t)$ by performing a time inversion and interchanging the drive phases. This symmetry will be an essential ingredient underlying the refocusing technique.

It is worth noticing that this symmetry arises specifically from the use of a bichromatic drive. The controlled-squeezing gate proposed in Ref.~\cite{PhysRevA.111.042606} uses a monochromatic drive in the implementation, and their Hamiltonian can be obtained from ours by setting one of the drive amplitudes ($\epsilon_0$ or $\epsilon_1$) to zero. In this case, within the RWA, each branch Hamiltonian contains either a static or a slowly oscillating term, but not both, and the symmetry discussed no longer holds. Consequently, the refocusing protocol described below cannot be applied, and to our knowledge no exact correction technique for the spurious effects induced by the time-dependent terms has been devised.



\hfill

\section{The refocusing technique}
\label{refocusing_section}
The refocusing technique, originally developed in the context of nuclear magnetic resonance \cite{Hahn1950}, provides a systematic way to suppress spurious effects arising from slowly varying time-dependent terms in the Hamiltonian. It consists of dividing the total evolution into two stages \cite{CarrPurcell1954, Sheldon2016}. During the first interval, $0<t<\tau$, unwanted contributions accumulate under a Hamiltonian that incorporates the relevant slowly varying time-dependent terms, such as $\hat H_I(t)$ in Eq.~\eqref{rwa_hamiltonian}. These contributions are canceled during the second interval, $\tau < t < 2\tau$, by evolving under its time-reversed counterpart, $\hat H_I(-t)$. The refocusing technique thus stands as a prominent candidate to cancel the spurious phase $\nu(t)$ appearing in Eq.~\eqref{state_evolved} during the implementation of the conditional-squeezing gate, as the symmetries of $\hat H_I(t)$ discussed above enable the required time reversal dynamics.

We show that the refocusing technique can be implemented by following the four-step protocol: (i) during the interval $0<t<\tau$, the qubit-resonator system evolves under the Hamiltonian $\hat{H}_I(t)$; (ii) at $t=\tau$, a $\pi$ pulse is applied on the qubit, exchanging $|0\rangle \leftrightarrow |1\rangle$, simultaneously with a permutation of the drive phases, $\varphi_0 \leftrightarrow \varphi_1$; (iii) during the interval $\tau<t<2\tau$, the system evolves with the transformed Hamiltonian: the combined action of the $\pi$ pulse and drive phases permutation reverses the sign of the time-dependent terms, so that the evolution is determined by the Hamiltonian $\hat{H}_I(-t)$; (iv) at $t=2\tau$, a second $\pi$ pulse is applied on the qubit.

\begin{figure}[!h]
\includegraphics[width=0.4\textwidth]{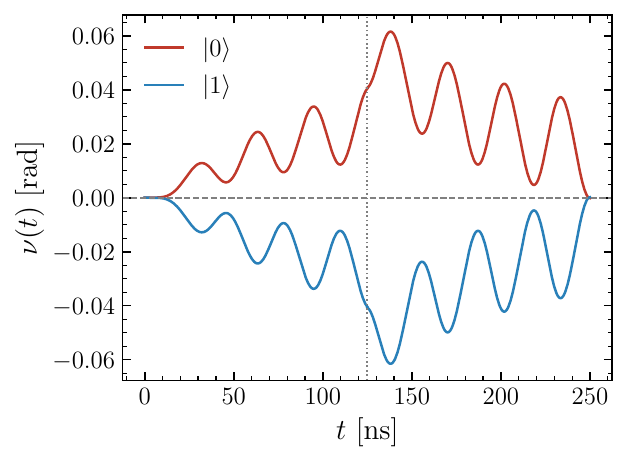}
\caption{\label{phase_reversion} Evolution of the induced phase $\nu(t)$ during the refocusing protocol of total duration $2\tau=250$~ns. Step (ii) of the protocol is applied at $t=125$ ns (dashed line). The evolution is shown for the qubit states $\ket{0}$ (red) and $\ket{1}$ (blue). }
\end{figure}

 
Numerical simulations of the protocol described above revealed that the refocusing suppresses the initial linear growth of $\nu(t)$, enabling phase cancellation at periodic time instants. Specifically, we found that effective cancellation occurs when the total protocol duration $2\tau$ is an integer multiple of the oscillation period $T=\pi/\chi=31.25$ ns of the time-dependent terms in $\hat{H}_I(t)$. The evolution of $\nu(t)$ for a protocol of $2\tau = 250$ ns (which corresponds to $8T$) is illustrated in Figure~\ref{phase_reversion}. Here the time-reversed stage (step (iii)) is initiated at $t = \tau=125$ ns (see dashed line), at which point the phase begins to decrease and reaches $\nu(2\tau)= 1.8\times10^{-8}$ rad. Since this cancellation mechanism extends to arbitrarily long protocol durations $2\tau=nT$, with $n$ an integer, the refocusing technique represents a significant improvement over the unrefocused evolution.

Beyond the phase cancellation, the quadrature oscillations can also be compensated by tuning the drive phases. As we discussed in Sec.~\ref{implementation}, the squeezing directions of the states oscillate around a value close to the target, so that they can be written as $\varphi_0/2+\delta\vartheta(t)$ and $\varphi_1/2-\delta\vartheta(t)$, where $\delta\vartheta$ is a time-dependent misalignment. By choosing the values of the drive phases appropriately, we can achieve $\delta\vartheta(2\tau)\approx0$. For a protocol duration of $2\tau=250$ ns, the choice $\varphi_0=0.0812$ rad and $\varphi_1=\pi-0.0812$ rad results in $\delta\vartheta(2\tau)\sim10^{-6}$ rad. Evidently, this misalignment can be made arbitrarily small by increasing the calibration precision of the driving phases. 

The restriction to discrete protocol durations does not limit the squeezing that can be achieved. Since the squeezing parameter at the end of the protocol is approximately $r\approx g_d nT/2$, a target value of $r$ can be reached by choosing the number of periods $n$ and tuning the coupling strength $g_d$, either through the flux drive amplitude, while keeping it within the weak-flux regime, or through the static bias $\Phi_s$ (see Sec.~\ref{implementation}). The only discrete quantity is the gate duration, in steps of $T$. Changing the ratio $g_d/\chi$ will require a recalibration of the drive phases.  We verified that, for $2\tau=9T$, varying $g_d$ to produce squeezing parameters between $r\approx1.25$ and $r\approx2.08$ and recalibrating the drive phases, the refocusing protocol yields residual phases $|\nu(2\tau)|\lesssim2\times10^{-8}$~rad.

Our results clearly indicate that the refocusing technique cancels the spurious effects induced by the slowly oscillating terms in the Hamiltonian $\hat H_I(t)$, and makes the resulting gate very close to an ideal conditional-squeezing gate $\text{CSq}(r,\varphi_0,\varphi_1)$. Below, we test the performance of the implemented gate with refocusing by numerically studying the fidelity of a simple encoding algorithm.

\hfill
\section{Encoding algorithm}
\label{encoding_section}
As a benchmark for the conditional-squeezing gate analyzed in the previous section, we consider an encoding algorithm that maps an arbitrary qubit state $\ket{\psi_{\text{Q}}}=\alpha\ket{0}+\beta\ket{1}$ into a resonator state $\ket{\psi_\text{R}}=\alpha \ket{0_L} +\beta \ket{1_L}$. The logical states $|0_L\rangle$ and $|1_L\rangle$ are two orthonormal resonator states, chosen as either $\ket{\chi_+}$ or $\ket{\chi_-}$, defined as even and odd superpositions of states squeezed along two orthogonal directions,

\begin{equation}
\label{chi_states}
    \ket{\chi_{\pm}}=\frac{1}{\sqrt{2}c_{\pm}}(\ket{r,0}\pm\ket{r,\pi}).
\end{equation}

\begin{widetext}
Here, $\ket{r,\vartheta}$ denotes a squeezed vacuum state with squeezing direction $\vartheta/2$ in phase space

\begin{equation}
\label{xi}
|r,\vartheta\rangle=\frac{1}{\sqrt{\cosh r}}
\sum_{k=0}^{\infty}\left(-e^{i\vartheta}\tanh r\right)^k\frac{\sqrt{(2k)!}}{2^k\,k!}\,|2k\rangle,
\end{equation}


\noindent and the normalization constants are $c_{\pm}=\sqrt{1\pm1/\sqrt{\cosh(2r)}}$. The states $\ket{\chi_\pm}$ can thus be expressed in terms of Fock states as

\begin{align}
|\chi_+\rangle&=\frac{1}{c_+\sqrt{2\cosh r}}\sum_{k=0}^{\infty}\frac{\sqrt{(4k)!}}{2^{2k-1}(2k)!}(\tanh r)^{2k}|4k\rangle, \\
|\chi_-\rangle&=\frac{-1}{c_-\sqrt{2\cosh r}}\sum_{k=0}^{\infty}\frac{\sqrt{(4k+2)!}}{2^{2k}(2k+1)!}(\tanh r)^{2k+1}|4k+2\rangle.
\end{align}
\end{widetext}

The virtue of this encoding is that errors induced by single-photon losses on $\ket{\psi_{\text{R}}}$ are detectable by means of a parity measurement on the resonator state. This follows from the fact that $\ket{\chi_+}$ and $\ket{\chi_-}$ have support only on Fock states with $4k$ and $4k+2$ photons, respectively:  a single photon loss maps each logical state to an orthogonal subspace with a different parity from the original one.




An algorithm that implements the above encoding was discussed in Ref.~\cite{PhysRevA.111.042606}, based on the controlled-squeezing gate. Since that gate squeezes the resonator for only one of the qubit states, the algorithm requires two applications of the gate, interleaved with $\pi$ rotations of the qubit. Here, we propose a generalization of that algorithm in which a single application of the conditional-squeezing gate suffices. We first describe it for an ideal conditional-squeezing gate, as defined in Eq.~\eqref{ideal_csq}. It consists of the following three steps: (i) a Hadamard gate is applied to the qubit, transforming $|0\rangle \rightarrow(|0\rangle+|1\rangle)/\sqrt{2}$ and $|1\rangle \rightarrow(|0\rangle-|1\rangle)/\sqrt{2}$; (ii) the ideal conditional-squeezing gate $\text{CSq}(r,0,\pi)$ is applied to the qubit--resonator system; (iii) a second Hadamard gate is applied to the qubit.


Starting from an arbitrary qubit state and the resonator in the vacuum, $\ket{\Psi_{\text{QR}}}=(\alpha\ket{0}+\beta\ket{1})\otimes \ket{0}$, the resulting state after applying the above algorithm is 

\begin{align}
\label{final_state}
|\Psi_{\mathrm{QR}}\rangle
=&\,\frac{1}{\sqrt{2}}|0\rangle\otimes
\left(
\alpha c_+ |\chi_+\rangle
+\beta c_- |\chi_-\rangle
\right)
 \nonumber \\ 
&+\, \frac{1}{\sqrt{2}}|1\rangle\otimes
\left(
\alpha c_- |\chi_-\rangle
+\beta c_+ |\chi_+\rangle
\right).
\end{align}

If we measure $\hat \sigma_z$ on the qubit, we obtain the results $\pm 1$ with probabilities $\text{Prob}({\pm}1)=1/2\left(1\pm P_z/\sqrt{\cosh{(2r)}}\right)$, respectively, where $P_z=|\alpha|^2-|\beta|^2$ is the $z$-component of the encoded state Bloch vector. For each outcome, the state of the resonator corresponds to


\begin{equation}
|\Psi^{\pm}_R\rangle = \frac{\alpha\, c_{\pm}|\chi_\pm\rangle + \beta\, c_\mp|\chi_\mp\rangle}{\sqrt{1 \pm \frac{P_z}{\sqrt{\cosh{(2r)}}}}}.
\end{equation}


Depending on the result of the $\hat \sigma_z$ measurement, we define the logical states $|0_L\rangle=\ket{\chi_\pm}$ and $|1_L\rangle=\ket{\chi_\mp}$, so that the encoded state is

\begin{equation}
\label{encoded_state}
|\Psi^{\pm}_{\text{enc}}\rangle = \frac{\alpha\, c_{\pm}|0_L\rangle + \beta\, c_\mp|1_L\rangle}{\sqrt{1 \pm \frac{P_z}{\sqrt{\cosh{(2r)}}}}}.
\end{equation}

Evidently, this algorithm implements a perfect encoding mapping the qubit state $\alpha\ket{0}+\beta\ket{1}$ onto the resonator state $\alpha \ket{0_L} +\beta \ket{1_L}$ only in the limit of infinite squeezing, where $c_\pm=1$. For finite squeezing, we can define the fidelity of the encoded state conditioned on each $\hat \sigma_z$ measurement outcome as $F_\pm=|\langle{\psi_{\text{R}}}|\Psi^{\pm}_\text{enc}\rangle|^2$. The average encoding fidelity of the algorithm $\bar{F}=P_+F_++P_-F_-$ can be written as

\begin{equation}
\label{theoretical_fidelity}
    \bar{F}=\frac{1}{2}(1+P_z^2)+\frac{1}{2}(1-P_z^2)\sqrt{1-\frac{1}{\cosh{(2r)}}}.
\end{equation}


This average fidelity depends on the azimuthal angle $\theta$ of the encoded state since $P_z=|P|\cos{\theta}$. In particular, it is maximal ($\bar{F}=1$) at the poles of the Bloch sphere (where $P_z=\pm1$) and minimal at the equator (where $P_z=0$). Moreover, the average fidelity at the equator increases with $r$. To achieve $\bar{F}\ge0.99$ for all initial qubit states, we need a squeezing of $r\gtrsim2$. Below, we present numerical results for the fidelity of the above encoding algorithm using the implementation of the conditional-squeezing gate developed in the previous sections. We divide the analysis into a unitary evolution (Sec.~\ref{unitary_evolution}) and one that includes Kerr-type nonlinearities and dissipation in the system (Sec.~\ref{open_system}), which inevitably reduce the fidelity.





\begin{figure}[!h]
\includegraphics[width=0.45\textwidth]{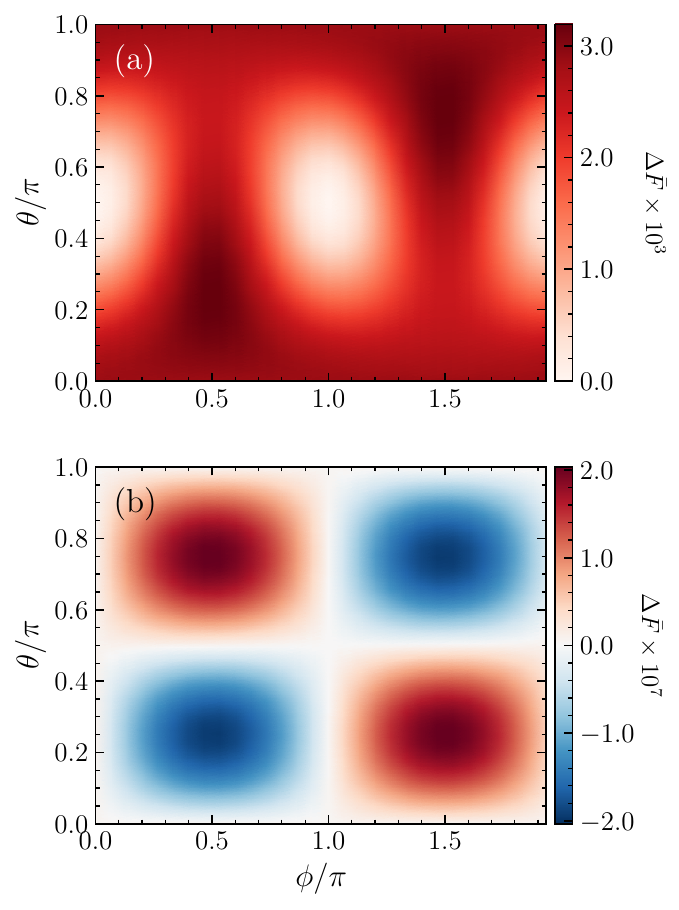}
\caption{\label{bloch_heatmap} Fidelity deviation $\Delta\bar{F}$ between an ideal encoding and that obtained using the implementation of the conditional-squeezing gate, across the Bloch sphere ($\theta$ and $\phi$ are the azimutal and polar angles, respectively). The results are shown for encoding algorithms  (a) with a non-compensated gate, and (b) with the gate compensated by the refocusing protocol. Note that the deviation in the top panel is four orders of magnitude larger than in the bottom one, as reflected by the different color scales.}
\end{figure}

\subsection{Unitary evolution}
\label{unitary_evolution}

We first study how the spurious effects induced by the time-dependent terms in the implementation of the conditional-squeezing gate impact the encoding fidelity, in the absence of dissipation and nonlinearities. For this purpose, we analyze the deviation $\Delta \bar{F}=\bar{F}-\bar{F}_{\text{num}}$, where $\bar{F}$ is the average fidelity in an ideal encoding, defined by Eq.~\eqref{theoretical_fidelity}, and $\bar{F}_{\text{num}}$ is the one obtained by solving the Schrödinger equation using the full Hamiltonian of Eq. \eqref{interaction_picture}, retaining all time-dependent terms, including the fast-oscillating ones neglected within the RWA. The $\pi$ pulses and Hadamard gates are treated as instantaneous operations. We numerically simulated the encoding algorithm using a total duration $2\tau=281.25$ ns, which corresponds to 9 periods of the slowly oscillating terms; as discussed in Sec.~\ref{refocusing_section}, the refocusing protocol cancels the induced phase only at discrete protocol durations of this kind. The algorithm generates squeezed states with $r\approx2.08$, for which the ideal encoding fidelity at the equator is $\bar{F}=0.9922$. 

First, we consider a non-compensated gate. Namely, we use a single application of $\hat H_I(t)$ in Eq. \eqref{interaction_picture} of duration $2\tau$, without applying any compensation method. We appropriately calibrated the drive phases so as to achieve the target directions at the end of the algorithm, isolating the effect of the induced phase in the encoding fidelity. In Figure~\ref{bloch_heatmap}(a) we show the fidelity deviation as a function of the polar ($\theta$) and azimuthal ($\phi$) angles of the Bloch sphere. In this case, the maximal fidelity deviation is $\Delta\bar{F}=3\times10^{-3}$.

The improvement achieved by the refocusing protocol is clearly seen in Figure~\ref{bloch_heatmap}(b), where the absolute fidelity deviation has a maximal value of $2\times10^{-7}$. In this case, $\Delta\bar F$ takes both positive and negative values across the Bloch sphere. Negative values do not indicate a fidelity larger than that of a perfect encoding: it is a consequence of the very small residual misalignments of the squeezed states at the end of the refocusing protocol ($\sim10^{-6}$~rad with appropriately calibrated drive phases), which can either increase or decrease the overlap with the target state (see Eq.~\eqref{encoded_state}) depending on the initial qubit state. This also explains why $\Delta\bar F$ depends on the azimuthal angle $\phi$, a feature absent in Eq.~\eqref{theoretical_fidelity}. Moreover, the success of the refocusing protocol validates the RWA used to obtain the Hamiltonian of Eq.~\eqref{rwa_hamiltonian}, on which the entire analysis of the refocusing-based compensation relies.

In the absence of nonlinearities and dissipation, these results do not depend on the particular value of the dispersive shift. Within the RWA, the interaction-picture Hamiltonian of Eq.~(3) is proportional to $g_d$ and depends on time only through $\chi t$, so that a simultaneous rescaling $\chi\to\lambda\chi$ and $g_d\to\lambda g_d$ transforms $\hat H_I(t)$ into $\lambda\hat H_I(\lambda t)$. The dynamics is therefore invariant up to a rescaling of time, $t\to t/\lambda$. This implies that, in a unitary evolution, the same squeezing parameter and fidelities are obtained for any value of $\chi$, provided that $g_d$ is scaled accordingly and the protocol duration is set to the same number of slow-oscillation periods. The fast-oscillating terms neglected within the RWA break this invariance only slightly, and their effect is negligible. For instance, using $\chi/2\pi=3$~MHz, the same results as above are obtained with $g_d/2\pi\approx0.45$~MHz and $2\tau=1.5~\mu$s. Within the model considered here, the choice of $\chi$ therefore affects the gate only through its duration, which determines the impact of dissipation and nonlinearities.

\subsection{The effects of nonlinearities and dissipation}
\label{open_system}

We now consider a realistic implementation of the conditional-squeezing gate, including the effects of Kerr-type nonlinearities and dissipation on the encoding fidelity. The Kerr nonlinearity arises from the quartic term in the expansion of the Josephson potential of the SQUID, which adds to the Hamiltonian of Eq. \eqref{full_hamiltonian} a correction of the form $\hat H_{\text{Kerr}}=(K/2) \hat a^{\dagger2}\hat a^2=(K/2)\hat n(\hat n-1)$, where $K$ is the nonlinearity strength and $\hat n=\hat a^\dagger \hat a$ is the photon number operator \cite{JJ, Nigg}. Here we only consider the nonlinearity of the SQUID. The dispersive coupling to the transmon also induces a self-Kerr nonlinearity of the resonator, which can be considerably larger than the SQUID-induced one and becomes relevant at the large photon numbers required by the encoding. Its origin, the conditions under which it can be neglected, and its scaling with the dispersive shift are discussed in Appendix \ref{apB}.


Before presenting the numerical results, it is instructive to estimate the impact of the Kerr term. For a squeezed state with mean photon number $\langle\hat n\rangle=\sinh^2 r$, $\hat H_{\rm Kerr}$ shifts the resonator frequency by an amount of order $K\langle\hat n\rangle$, which detunes the parametric drive and tilts the squeezing direction by an angle $\delta\sim K\langle\hat n\rangle/g_d$. The relevant scale for this tilt is the angular width of the squeezed state, of order $1/\sinh(2r)$: the overlap between two squeezed vacuum states whose squeezing directions differ by $\delta$ is approximately $1-\sinh^2(2r)\,\delta^2/2$. The Kerr-induced fidelity loss is therefore of order $[\sinh(2r)K\langle\hat n\rangle/g_d]^2$. For $K/2\pi=100$~Hz, $g_d/2\pi=2.39$~MHz and $r=2.08$, we obtain $\delta\sim6\times10^{-4}$ and a fidelity loss of order $4\times10^{-4}$. Since $g_d\propto\chi$ for a fixed squeezing parameter and a fixed number of periods, this contribution grows as $\chi^{-2}$ and becomes relevant for smaller dispersive shifts, as we will show below. The value $K/2\pi=100$~Hz is rather optimistic, as typical Kerr strengths in SQUID-terminated resonators can reach 10 kHz \cite{PhysRevB.87.184501, Krantz2013}. However, we note that the Kerr effect could be further suppressed by using instead a SNAIL-terminated resonator, for which a Kerr-free regime can be achieved by appropriately tuning the external flux \cite{Frattini2018, Sivak2019, Frattini2024, Lu2023}.

Dissipation is modeled through a Lindblad master equation describing the coupling of the qubit-resonator system to a thermal bath. We account for photon loss and thermal excitation in the resonator, with decay time $\tau_r$, and for relaxation, thermal excitation, and pure dephasing of the qubit, with timescales $\tau_q$ (for the first two) and $\tau_\phi$ (for dephasing). In the interaction picture, the density matrix of the qubit-resonator system evolves according to the master equation

\begin{align}
\dot{\hat\rho}(t)={}&-i\big[\hat H_I(t)+\hat H_{\text{Kerr}},\hat\rho(t)\big]\nonumber\\
&+\sum_{i=1}^{5}\kappa_i\Big(\hat L_i(t)\hat\rho(t)\hat L_i^\dagger(t)-\tfrac12\hat L_i^\dagger(t)\hat L_i(t)\hat\rho(t)\nonumber\\
&\qquad\qquad-\tfrac12\hat\rho(t)\hat L_i^\dagger(t)\hat L_i(t)\Big).
\label{master_equation}
\end{align}

The Lindblad operators are $\hat L_1=\hat a$, $\hat L_2=\hat a^{\dagger}$, $\hat L_3=\hat\sigma_-$, $\hat L_4=\hat\sigma_+$, and $\hat L_5=\hat\sigma_z$, with $\hat\sigma_-=|1\rangle\langle0|$ and rates $\kappa_1=(n_r+1)/\tau_r$, $\kappa_2=n_r/\tau_r$, $\kappa_3=(n_q+1)/\tau_q$, $\kappa_4=n_q/\tau_q$, and $\kappa_5=1/(2\tau_\phi)$. Here, $n_r=1/[\exp(\omega/k_BT)-1]$ and $n_q=1/[\exp(\omega_q/k_BT)-1]$ are the thermal occupation numbers of the resonator and qubit environments, respectively. In the interaction picture, the Lindblad operators become time-dependent through $\hat a(t)=\hat a\,e^{-i(\omega+\chi\hat\sigma_z/2)t}$, $\hat a^\dagger(t)=[\hat a(t)]^\dagger$, $\hat\sigma_-(t)=\hat\sigma_-e^{-i(\omega_q+\chi\hat n)t}$, $\hat\sigma_+(t)=\hat\sigma_-^\dagger(t)$, and $\hat\sigma_z(t)=\hat\sigma_z$. The resulting time-dependent phases, which depend on the qubit state and on the photon number, are taken into account in our simulations (see Appendix~\ref{app:blockME}).

\begin{figure}[h!]
\includegraphics[width=\columnwidth]{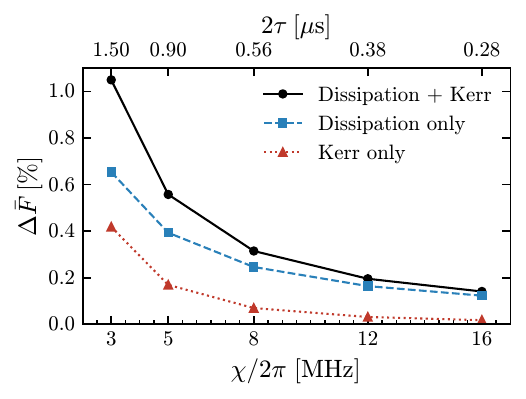}
\caption{Fidelity deviation $\Delta\bar F$ of the encoding algorithm for an initial qubit state at the equator of the Bloch sphere ($\theta=\pi/2$, $\phi=\pi/2$), as a function of the dispersive shift $\chi$ (bottom axis) and of the corresponding gate duration $2\tau$ (top axis). The drive strength is scaled as $g_d\propto\chi$, so that $r\approx2.08$ in all cases. Shown are the total loss (dissipation and Kerr nonlinearity, $K/2\pi=100$~Hz) and the separate contributions of dissipation ($K=0$) and of the Kerr nonlinearity (no dissipation). The results for states at the poles are nearly identical.}
\label{fig:chi_sweep}
\end{figure}

 For the qubit dissipation channels ($i=3,4,5$), the expected fidelity decay is of order $\kappa_i 2\tau$  when $\kappa_i$ is small. For highly optimistic but achievable transmon coherence times $\tau_q=\tau_\phi=1$~ms \cite{Bland2025, PhysRevLett.130.267001} and a bath temperature $T=60$ mK, these corrections amount to a few $10^{-4}$ for $2\tau=281.25$~ns. Photon losses in the resonator are expected to be the dominant contribution to fidelity loss due to the large number of photons in the squeezed states. For this reason, we consider a resonator lifetime $\tau_r=1$~ms, which is optimistic for a SQUID-terminated resonator but consistent with recent results for on-chip superconducting resonators~\cite{Ganjam2024}. The corresponding fidelity loss for photon loss can be estimated as the probability of a photon-loss event during the gate, $\kappa_1\int\langle\hat n(t)\rangle\,dt$. With $\langle\hat n(t)\rangle=\sinh^2r(t)$ and $r(t)$ growing linearly up to its final value $r$, this gives, for $2\tau=281.25$~ns, a fidelity loss of approximately $1\times10^{-3}$.

To obtain the encoding fidelity including all these effects, we solved the master equation of Eq.~\eqref{master_equation} by direct numerical integration, using a decomposition of the density matrix into qubit blocks described in Appendix~\ref{app:blockME}. We used a truncated Hilbert space for the resonator of dimension $N=500$, for which the results are converged to $10^{-6}$. The drive phases were calibrated in the presence of the Kerr term, as would be done in an experiment.

For $\chi/2\pi=16$~MHz and $2\tau=281.25$~ns, we found that the two states at the poles of the Bloch sphere ($\theta=0,\pi$) have a fidelity of $0.9986$, whereas the states at the equator ($\theta=\pi/2$, with $\phi=0$ and $\phi=\pi/2$) have fidelities of $0.9909$ and $0.9908$, respectively ($\bar F=0.9922$ for an ideal encoding). These results represent fidelity deviations below 0.15\%, and encoding fidelities above 0.99 for arbitrary initial qubit states. When only dissipation is included, the fidelity loss is 0.12\%, in agreement with the estimates above, so that photon loss in the resonator is the dominant contribution.

For smaller dispersive shifts, the same encoding is obtained with a proportionally weaker drive and a longer gate, $2\tau\propto1/\chi$, so that dissipation and nonlinearities act for a longer time. Figure~\ref{fig:chi_sweep} shows the fidelity deviation $\Delta \bar{F}$ for an initial qubit state at the equator of the Bloch sphere ($\theta=\pi/2$, $\phi=\pi/2$) as a function of $\chi$, for $\chi/2\pi$ between 3 and 16~MHz (gate durations between $1.5~\mu$s and $281$~ns), with the drive phases calibrated in the presence of the Kerr term for each value of $\chi$. We separate the contributions of dissipation and of the Kerr nonlinearity. The dissipative loss grows linearly with the gate duration, that is, as $1/\chi$, from $0.12\%$ at $\chi/2\pi=16$~MHz to $0.65\%$ at 3~MHz. The Kerr contribution (for $K/2\pi=100$~Hz), in contrast, grows approximately quadratically with the gate duration, as anticipated by the estimate above, from $0.02\%$ at $\chi/2\pi=16$~MHz to $0.42\%$ at 3~MHz, so that dissipation remains the dominant contribution in the whole range. Both contributions are approximately additive. For the parameters considered here, the equatorial fidelity remains above $0.99$ for $\chi/2\pi\gtrsim12$~MHz; for smaller dispersive shifts, maintaining this performance requires proportionally longer resonator lifetimes and, to a lesser extent, a smaller Kerr nonlinearity.

\subsection{Comparison with the controlled-squeezing gate}
\label{comparison_section}

To quantify the improvement over Ref.~\cite{PhysRevA.111.042606}, we simulated the encoding algorithm proposed there, which requires two applications of the controlled-squeezing gate, interleaved with $\pi$ pulses on the qubit. We performed the simulations with the same parameters as in the previous sections. For the same driving strength and approximately the same final squeezing, each gate application lasts $2\tau$, so that the total duration of the algorithm is twice that of the one based on the conditional gate. In the controlled-squeezing gate, the branch that is not resonantly driven acquires a residual squeezing produced by the slowly oscillating terms, which in Ref.~\cite{PhysRevA.111.042606} is approximated by a free rotation in phase space and compensated accordingly. We used this compensation in our simulations of the algorithm of Ref.~\cite{PhysRevA.111.042606}. This inaccurate compensation of the time-dependent terms reduces the fidelity, even in the absence of nonlinearities and dissipation, by $4.7\%$ on average over the Bloch sphere ($7.8\%$ for the worst initial state) for the $200$~ns gate duration used in Ref.~\cite{PhysRevA.111.042606} ($r\approx1.5$). The fidelity deviation becomes smaller, $2\times10^{-4}$ on average ($3\times10^{-4}$ in the worst case), only when the duration of each application is an integer multiple of $T$, as for $2\tau=9T$, while it remains below $3\times10^{-7}$ for the conditional-squeezing gate with refocusing.

\begin{figure}[h!]
\includegraphics[width=\columnwidth]{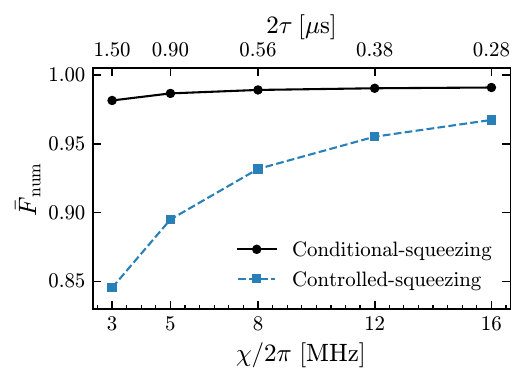}
\caption{Numerical fidelity $\bar F_{\text{num}}$ of the encoding algorithms based on the conditional-squeezing gate (this work) and on the controlled-squeezing gate of Ref.~\cite{PhysRevA.111.042606}, for an initial qubit state at the equator of the Bloch sphere ($\theta=\pi/2$, $\phi=\pi/2$), as a function of $\chi$ (bottom axis), including dissipation and the Kerr nonlinearity ($K/2\pi=100$~Hz). The top axis indicates the total duration $2\tau=9T$ of the algorithm based on the conditional-squeezing gate; the algorithm based on the controlled-squeezing gate consists of two applications of duration $9T$ each, and thus lasts twice as long. Both algorithms use the same drive strength per tone and reach the same squeezing parameter, $r\approx 2.08$.}
\label{fig:comparison}
\end{figure}

In Figure~\ref{fig:comparison}, we compare the performance of both gates in the encoding algorithm, including dissipation and the Kerr nonlinearity. The algorithm based on the conditional-squeezing gate has a total duration $2\tau=9T$, whereas the one based on the controlled-squeezing gate has a total duration $18T$, divided into two applications of $9T$ each. For $\chi/2\pi=16$~MHz, the equatorial fidelity of the algorithm of Ref.~\cite{PhysRevA.111.042606} is $0.967$, compared with $0.991$ for the conditional-squeezing gate; for $\chi/2\pi=3$~MHz, these values are $0.845$ and $0.981$, respectively, with similar differences at the poles. The main origin of this difference is that, in the algorithm of Ref.~\cite{PhysRevA.111.042606}, the state squeezed during the first application remains idle during the entire second application. In the absence of the drive, the Kerr term distorts this state through a photon-number-dependent rotation, which, unlike in our protocol, becomes the dominant source of fidelity loss: in the absence of dissipation, it reduces the equatorial fidelity by $2.2\%$ at $\chi/2\pi=16$~MHz and by $13.6\%$ at 3~MHz, compared with $0.02\%$ and $0.42\%$ for the conditional gate. In the latter, instead, both branches are squeezed simultaneously, the drive remains on during the whole gate, which limits the Kerr-induced tilt of the squeezing direction (see Sec.~\ref{open_system}), and the photon number reaches its maximum only at the end of the gate. As a result, this advantage increases as $\chi$ decreases.

\section{The two-mode conditional-squeezing gate}
\label{tms_section}
We now extend the single-mode conditional-squeezing gate developed in the previous sections to a conditional-squeezing gate acting on two resonator modes. Two-mode squeezed states (TMS states) are entangled Gaussian states that represent the continuous-variable analog of Bell states \cite{CavesSchumaker1985, Weedbrook2012} in the limit of infinite squeezing, and are characterized by Einstein-Podolsky-Rosen (EPR) correlations between the quadratures of the two modes. TMS states represent a key resource in continuous-variable quantum information processing, serving as the entangled channel for quantum teleportation~\cite{Takeda2013}, dense coding~\cite{BraunsteinKimble2000,Li2002}, and measurement-based quantum computation~\cite{PhysRevLett.97.110501,Asavanant2019, Larsen2019}.

TMS states are most naturally described in terms of the EPR-like variables $\hat x_\pm$ and $\hat p_\pm$, defined as symmetric and antisymmetric combinations of the position $\hat{x}_{1,2}$ and momentum $\hat{p}_{1,2}$ quadratures of each mode

\begin{equation}
\label{tms_quadratures}
\hat{x}_{\pm} = \frac{1}{\sqrt{2}}(\hat{x}_1 \pm \hat{x}_2), \qquad \hat{p}_{\pm} = \frac{1}{\sqrt{2}}(\hat{p}_1 \pm \hat{p}_2).
\end{equation}

In an ideal TMS state, the quadratures $\hat x_-$ (that represents the relative position) and $\hat p_+$ (that represents the total momentum) are simultaneously squeezed, while the conjugate combinations $\hat x_+$ and $\hat p_-$ are correspondingly anti-squeezed. The TMS vacuum state is obtained by the action of the operator $\hat{S}_2(r, \vartheta)$ on the two-mode vacuum state $\ket{0,0}$, where $\hat{S}_2(r, \vartheta)$ is
 
\begin{equation}
\hat{S}_2(r, \vartheta) = \exp\left[r\left(\hat{a}_1 \hat{a}_2\, e^{-i\vartheta} - \hat{a}_1^\dagger \hat{a}_2^\dagger\, e^{i\vartheta}\right)\right].
\end{equation}

As usual, $r$ represents the squeezing parameter and $\vartheta$ the orientation of the correlations in phase space. The TMS vacuum state, $|r,\vartheta\rangle_2=\hat S_2(r,\vartheta)|0,0\rangle$, can be written in terms of Fock states as

\begin{equation}
\label{tms_operator}
|r,\vartheta\rangle_2 = \frac{1}{\cosh r}\sum_{k=0}^{\infty}(-e^{i\vartheta}\tanh r)^k|k,k\rangle,
\end{equation}

\noindent which is a superposition of Fock states containing the same photon number in both modes. 


The generalization of the superposition states $\ket{\chi_\pm}$ for the two-mode case results $\ket{\chi_{\pm}}_2=\frac{1}{\sqrt{2}\mathcal{N}_\pm}(\ket{r,0}_2\pm\ket{r,\pi}_2)$, where $\mathcal{N}_\pm$ is a normalization constant (see below). When written in term of Fock states, $\ket{\chi_{\pm}}_2$ are

\begin{align}
\label{chi_mas_tms}
|\chi_+\rangle_2 &= \frac{\sqrt{2}}{\mathcal{N}_+ \cosh r}\sum_{k=0}^{\infty}\left(e^{i\vartheta}\tanh r\right)^{2k}|2k,2k\rangle, \\
|\chi_-\rangle_2 &=  \frac{\sqrt{2}}{\mathcal{N}_-\cosh r}\sum_{k=0}^{\infty}\left(e^{i\vartheta}\tanh r\right)^{2k+1}|2k+1,2k+1\rangle,
\label{chi_menos_tms}
\end{align}

\noindent with $\mathcal N_\pm=\sqrt{1\pm1/\cosh(2r)}$. It is evident that $|\chi_+\rangle_2$ ($|\chi_-\rangle_2$) has support only on Fock states pairs with even (odd) photon number. Thus, a photon loss will induce a change in the photon number parity in either mode. A possible way of detecting this loss is by measuring the photon number difference $\hat n_1-\hat n_2$ between the two modes, which is $+1$ $(-1)$ if the photon loss occurs on the second (first) mode, both for $|\chi_+\rangle_2$ and $|\chi_-\rangle_2$.

\hfill

\subsection{Implementation}
\label{subsection_tms}
The two-mode conditional-squeezing gate can be implemented in the setup of Figure~\ref{circuit} by considering two modes of the SQUID-terminated resonator, each of which is also coupled to the qubit. The Hamiltonian in this case can be written as 

\begin{equation}
\label{hamiltonian_tms}
    \hat{H}_{\text{tm}}(t)=\frac{\omega_q}{2}\hat \sigma_z+\hat H_1(t)+\hat H_2(t)+\hat{H}_{\text{int}}(t). 
\end{equation} 

Here, $\hat H_1(t)$ and $\hat H_2(t)$ are the Hamiltonians for each of the two modes stored in the SQUID-terminated resonator, defined by the last three terms of Eq.~\eqref{full_hamiltonian}. The contribution $\hat H_i(t)$ ($i=1,2$) depends on the bare resonator mode frequency $\omega_{i}$, the dispersive coupling constant $\chi_{i}$ and the coupling to the driving field $\bar{g}_{d,i}$. In turn, the interaction term between the modes $\hat{H}_{\text{int}}(t)$ can be written on the form

\begin{align}
\label{interaction_hamiltonian_tms}
    \hat H_{\text{int}}(t)=i\Gamma\delta\dot\Phi(t)\Big[(\omega_1-\omega_2)(\hat a_1\hat a_2-\hat a^{\dagger}_1\hat a^{\dagger}_2) \nonumber \\
    +(\omega_1+\omega_2)(\hat a_1\hat a^{\dagger}_2-\hat a^{\dagger}_1\hat a_2)\Big],
\end{align}

\noindent where $\Gamma$ is a coupling constant and $\delta\dot\Phi(t)$ is the time-derivative of the external flux. A detailed derivation of $\hat H_{\text{int}}(t)$ is provided in Appendix \ref{apA}. 

The Hamiltonian of Eq.~\eqref{hamiltonian_tms} describes two interacting harmonic modes whose frequency depends both on the state of the qubit (through the dispersive coupling) and on time (through the external flux). We define the state- and time-dependent frequencies as $\tilde{\omega}_{i,\pm} = \omega_{i,\pm} + \bar{g}_{d,i}\delta\Phi(t)$, where the constant term is $\omega_{i,\pm}=\omega_i\pm \frac{\chi_i}{2}$. In the previous expression, $+$ ($-$) corresponds to the $|0\rangle$ $(|1\rangle)$ state of the qubit.

Using a bichromatic drive of the form $\delta \Phi(t)=\epsilon_0\sin{(\omega_{d,0}t-\varphi_0)}+ \epsilon_1\sin{(\omega_{d,1}t-\varphi_1)}$, we excite state-dependent parametric resonance to generate selective two-mode squeezing. To do this, we choose the driving frequencies as $\omega_{d,0}=\omega_{1,+}+\omega_{2,+}$ and $\omega_{d,1}=\omega_{1,-}+\omega_{2,-}$, and write the Hamiltonian of Eq.~\eqref{hamiltonian_tms} in the interaction picture with respect to its static terms: $\hat{H}_{\text{tm},I}(t)=\hat H_{1,I}(t)+\hat H_{2,I}(t)+\hat H_{\text{int},I}(t)$. Applying the same procedure as in Sec.~\ref{implementation}, one can obtain the explicit form of $\hat H_{1,I}(t)$ and $\hat H_{2,I}(t)$. We note, however, that this choice of drive frequencies never excites parametric resonance of each mode independently. In fact, it is simple to see that these Hamiltonians contain purely oscillatory terms, of which the lowest frequencies are of the order $|\omega_2-\omega_1|=2\pi\times4$ GHz. For this reason, both $\hat H_{1,I}(t)$ and $\hat H_{2,I}(t)$ become negligible within the RWA.

The interaction Hamiltonian $\hat H_{\text{int},I}(t)$, on the other hand, reads as 

\begin{widetext}
\begin{align}
\hat H_{\text{int},I}(t)={}&i\Gamma\,\delta\dot\Phi(t)\,|0\rangle\langle0|\otimes\Big[(\omega_1-\omega_2)\big(\hat a_1\hat a_2e^{-i(\omega_{1,+}+\omega_{2,+})t}-\hat a^{\dagger}_1\hat a^{\dagger}_2e^{i(\omega_{1,+}+\omega_{2,+})t}\big)\nonumber\\
&\qquad\qquad\qquad\quad+(\omega_1+\omega_2)\big(\hat a_1\hat a^{\dagger}_2e^{-i(\omega_{1,+}-\omega_{2,+})t}-\hat a^{\dagger}_1\hat a_2e^{i(\omega_{1,+}-\omega_{2,+})t}\big)\Big]\nonumber\\
&+i\Gamma\,\delta\dot\Phi(t)\,|1\rangle\langle1|\otimes\Big[(\omega_1-\omega_2)\big(\hat a_1\hat a_2e^{-i(\omega_{1,-}+\omega_{2,-})t}-\hat a^{\dagger}_1\hat a^{\dagger}_2e^{i(\omega_{1,-}+\omega_{2,-})t}\big)\nonumber\\
&\qquad\qquad\qquad\quad+(\omega_1+\omega_2)\big(\hat a_1\hat a^{\dagger}_2e^{-i(\omega_{1,-}-\omega_{2,-})t}-\hat a^{\dagger}_1\hat a_2e^{i(\omega_{1,-}-\omega_{2,-})t}\big)\Big].
\label{interaction_picture_hamiltonian_tms}
\end{align}

We can write each branch of $\hat H_{\text{int},I}(t)$ (i.e., the terms proportional to $|0\rangle\langle0|$ and $|1\rangle\langle1|$) as the sum of sixteen oscillatory terms. Two of them, proportional to the operators $\hat a_1\hat a_2$ and $\hat a^\dagger_1\hat a^\dagger_2$, are static terms which arise because $\delta\Phi(t)$ satisfies the parametric resonance condition detailed above. The same operators also contribute the slowest oscillatory terms of frequencies $|\omega_{1,\pm}-\omega_{2,\mp}-\omega_{1,\mp}+\omega_{2,\pm}|=\chi_1+\chi_2\sim (2\pi)\times 10$ MHz, for $\chi_1,\chi_2\sim(2\pi)\times 10$ MHz. The remaining twelve terms in each branch oscillate at much higher frequencies, which differ from the previous ones by terms proportional to $\omega_{1,\pm}$ and $\omega_{2,\pm}$. Therefore, all of these last twelve terms can be neglected within the RWA. Under the condition $\epsilon_0\omega_{d,0}=\epsilon_1\omega_{d,1}$ on the drive amplitudes, the interaction picture Hamiltonian, retaining only the static and slowly oscillating terms, reads as (we define $\gamma=\Gamma(\omega_1-\omega_2)\epsilon_0\omega_{d,0}/2$)

\begin{align}
\hat{H}_{\text{tm},I}(t) & \approx|0\rangle\langle 0| \otimes \left(-i\gamma\right) \left\{ \hat{a}_1\hat{a}_2 e^{-i\varphi_0} - \hat{a}_1^\dagger\hat{a}_2^\dagger e^{i\varphi_0} + \hat{a}_1\hat{a}_2 e^{-i(2(\chi_1+\chi_2)t+\varphi_1)} - \hat{a}_1^\dagger\hat{a}_2^\dagger e^{i(2(\chi_1+\chi_2)t+\varphi_1)} \right\}  \nonumber  \\ 
&+ |1\rangle\langle 1| \otimes \left(-i\gamma\right) \left\{ \hat{a}_1\hat{a}_2 e^{-i\varphi_1} - \hat{a}_1^\dagger\hat{a}_2^\dagger e^{i\varphi_1} + \hat{a}_1\hat{a}_2 e^{-i(-2(\chi_1+\chi_2)t+\varphi_0)} - \hat{a}_1^\dagger\hat{a}_2^\dagger e^{i(-2(\chi_1+\chi_2)t+\varphi_0)} \right\}.
\label{approximate_hamiltonian_tms}
\end{align}
\end{widetext}

The structure of $\hat{H}_{\text{tm},I}(t)$ in Eq.~\eqref{approximate_hamiltonian_tms} is reminiscent of that found for $\hat H_I(t)$ in Eq.~\eqref{rwa_hamiltonian} in the single-mode case, with each branch Hamiltonian containing a static and a slowly oscillating contribution. Retaining only the static terms, the evolution operator associated with $\hat{H}_{\text{tm},I}(t)$ reduces to an ideal two-mode conditional-squeezing operator, which we denote as $\text{CSq}_{\text{2}}(r,\varphi_0,\varphi_1)$

\begin{equation}
\label{ideal_csq_two_mode}
    \text{CSq}_{\text{2}}(r,\varphi_0,\varphi_1)=|0\rangle\langle 0| \otimes \hat S_2(r,\varphi_0)+ |1\rangle\langle 1| \otimes \hat S_2(r,\varphi_1),
\end{equation}

\noindent where the magnitude of the squeezing parameter is $|r|=\gamma t$. The slowly oscillating terms in Eq.~\eqref{approximate_hamiltonian_tms} generate the three main effects already studied for the single-mode gate: (i) a small harmonic modulation of the squeezing parameter, (ii) oscillations of the squeezing direction of the two-mode squeezed states $\ket{\xi_0(t)}_2$ and $\ket{\xi_1(t)}_2$ associated to each branch, and (iii) a time-dependent phase  $\nu_2(t)$ as in Eq.~\eqref{state_evolved}. It stands to reason that the refocusing protocol is directly applicable to suppress these spurious effects.

We numerically simulated the four-step refocusing protocol presented in Sec.~\ref{refocusing_section} for a unitary evolution under $\hat H_{\text{tm},I}(t)$. For the second resonator mode, we used a mode frequency $\omega_2/2\pi=10$~GHz and a dispersive shift $\chi_2/2\pi=12$~MHz (the parameters of the first mode are the same as in Sec.~\ref{implementation}), and an effective coupling between the modes $\gamma/2\pi=1.19$~MHz, which sets the two-mode squeezing rate $dr/dt=\gamma$, the same as in the single-mode case. All these values can be achieved with current experimental capabilities~\cite{Wulschner2016}. With these parameters, the oscillation period of the time-dependent terms is $T_2=2\pi/(\chi_1+\chi_2)\approx35.7$~ns, very similar to the one obtained in the single-mode case. We performed the numerical simulations with a customized implementation that integrates the Schrödinger equation with a fourth-order Runge--Kutta method, using a Hilbert space dimension of $N=250$ for each mode.

We considered a total protocol duration of $2\tau=7T_2=250$~ns, for which $r=1.83$. The drive phases were set to $\varphi_0=-0.14280$~rad and $\varphi_1=\pi-0.14280$~rad, resulting in residual misalignments of $\sim10^{-6}$~rad at the end of the protocol with respect to the target directions ($0$ and $\pi/2$ for the qubit states $\ket{0}$ and $\ket{1}$, respectively). We confirmed that the refocusing technique cancels the induced phase $\nu_2(t)$ at the end of the protocol ($\nu_2(2\tau)\sim10^{-5}$~rad), so that the resulting gate is very close to the ideal two-mode conditional-squeezing gate $\text{CSq}_2(r,\varphi_0,\varphi_1)$. As in the single-mode case, since $\hat H_{\text{tm},I}(t)$ is proportional to $\gamma$ and depends on time only through $(\chi_1+\chi_2)t$, the same results are obtained for smaller dispersive shifts by scaling $\gamma$ accordingly, at the cost of proportionally longer gates. Below, we test its performance in a generalization to two modes of the encoding algorithm of Sec.~\ref{encoding_section}.

\subsection{Encoding algorithm}
We use the encoding algorithm presented in Sec.~\ref{encoding_section} as a benchmark for the two-mode conditional-squeezing gate. The three-step algorithm remains unchanged, except that the two-mode gate is now used in step (ii) instead of the single-mode one. Correspondingly, we choose the logical states $\ket{0_L}$ and $\ket{1_L}$ as either $\ket{\chi_+}_2$ or $\ket{\chi_-}_2$ (depending on the outcome of the qubit measurement), see Eqs.~\eqref{chi_mas_tms} and~\eqref{chi_menos_tms}.  Thus, the resulting qubit-resonator state after applying the algorithm is 

\begin{align}
\label{final_state}
|\Psi_{\mathrm{QR}}\rangle
=&\,\frac{1}{\sqrt{2}}|0\rangle\otimes
\left(
\alpha\mathcal{N}_+ |\chi_+\rangle_2
+\beta\mathcal{N}_- |\chi_-\rangle_2
\right) \nonumber
\\ 
&+\, \frac{1}{\sqrt{2}}|1\rangle\otimes
\left(
\alpha \mathcal{N}_- |\chi_-\rangle_2
+\beta \mathcal{N}_+ |\chi_+\rangle_2
\right),
\end{align}

\noindent where we recall $\mathcal N_\pm=\sqrt{1\pm1/\cosh(2r)}$. The average fidelity in an ideal encoding for the two-mode gate, which we denote as $\bar{F}_2$, can be written as

\begin{equation}
\label{fidelities_TMS}
\bar{F}_2 = \frac{1}{2}(1+P_z^2) + \frac{1}{2}(1-P_z^2)\sqrt{1 - \frac{1}{\cosh^2(2r)}}.
\end{equation}

A remarkable feature of $\bar F_2$ is that it approaches unity faster with $r$ than $\bar F$ for the single-mode gate (for $P_z\neq\pm1$). This is because the two-mode squeezed states composing the logical states have a smaller overlap, ${}_2\langle r,0|r,\pi\rangle_2=1/\cosh(2r)$, compared with $1/\sqrt{\cosh(2r)}$ in the single-mode case, which is a direct consequence of the entangled nature of TMS states. As a result, higher fidelities are reached at smaller squeezing parameters. This improvement is most evident for equatorial states ($P_z=0$): at $r=1.8$, $\bar F_2\approx0.99925$, compared with $\bar F\approx0.9862$.

We now present numerical results for an encoding algorithm of total duration $2\tau=7T_2=250$~ns, for which the ideal equatorial fidelity is approximately $0.99935$. Using the two-mode conditional-squeezing gate with refocusing, in a unitary evolution under $\hat{H}_{\text{tm},I}(t)$, the fidelity deviation $\Delta\bar{F}_2$ (defined as in Sec.~\ref{encoding_section}) remains below $4\times10^{-7}$ for all initial qubit states. In contrast, the fidelity deviation obtained without refocusing is significantly higher, of order $10^{-2}$ for states at the equator of the Bloch sphere. This clearly demonstrates that the refocusing technique is highly effective in this case.

To account for the nonlinearities induced by the Josephson junctions of the SQUID, which are shared between the two modes, we use the black-box quantization approach of Ref.~\cite{Nigg2012}. This approach relates the self-Kerr strengths of the two modes, $K_1$ and $K_2$, through the circuit geometry, and fixes the cross-Kerr strength as $K_{12}=2\sqrt{K_1K_2}$ (see Appendix~\ref{apA}). Assuming an optimistic value $K_1/2\pi=100$~Hz for the first mode, as in Sec.~\ref{open_system}, we obtain $K_2/2\pi=226.1$~Hz and $K_{12}/2\pi=300.7$~Hz. Numerical simulations of the encoding algorithm including these small strength Kerr nonlinearities show only a small decrease of the fidelity, $\Delta\bar F_2\sim10^{-5}$.

We modeled the dissipative effects using a Lindblad master equation that generalizes Eq.~\eqref{master_equation} to the two-mode resonator. Specifically, we include photon loss and thermal excitation in the second mode of the resonator through the Lindblad operators $\hat L_6=\hat a_2$ and $\hat L_7=\hat a^{\dagger}_2$, with rates $\kappa_6=(n_{r,2}+1)/\tau_{r,2}$ and $\kappa_7=n_{r,2}/\tau_{r,2}$, where $n_{r,2}=1/[\exp(\omega_2/k_BT)-1]$ is the thermal occupation number of the second resonator mode. We use the same optimistic decay time $\tau_{r,1}=1$~ms for the first mode as in Sec.~\ref{open_system} and, assuming the same quality factor for the two modes, set $\tau_{r,2}=\tau_{r,1}\omega_1/\omega_2=0.6$~ms for the second mode. 

In this case, we solved the master equation using quantum trajectories, truncated at one quantum jump, which is a very good approximation for the rates considered. We computed the fidelity loss for four cardinal states of the Bloch sphere: the two states at the poles ($\theta=0,\pi$) and two states at the equator ($\theta=\pi/2$, with $\phi=0$ and $\pi/2$). In all cases, the fidelity loss is approximately $0.17\%$, dominated by photon loss in the resonator, resulting in a minimum fidelity at the equator of $0.9977$. Compared with the single-mode gate, the two-mode scheme thus reaches a higher encoding fidelity for all initial qubit states ($0.9977$ versus $0.9908$ at the equator), with a smaller squeezing parameter and a shorter gate time, while the fidelities at the poles are comparable. As discussed above, this improvement is a consequence of the entangled nature of TMS states.

\hfill
\section{Conclusions}
\label{conclusion}
We presented an implementation of the single- and two-mode conditional-squeezing gate, in which a resonator state is squeezed along a direction determined by the state of a control qubit. The physical mechanism underlying the conditional-squeezing operation is the excitation of parametric resonance dependent on the state of the qubit, enabled by the dispersive qubit–resonator interaction and a bichromatic flux drive acting on the SQUID that terminates the resonator. This bichromatic flux drive introduces slowly oscillating terms in the Hamiltonian, which cause deviations from the ideal conditional-squeezing gate. We showed that a refocusing protocol, based on a time-reversal symmetry of the Hamiltonian, cancels these effects and is essential for building a high-fidelity gate.

As a benchmark, we applied the conditional-squeezing gate in a quantum algorithm that encodes an arbitrary qubit state into even and odd superpositions of squeezed vacuum states along orthogonal directions, $|\chi_\pm\rangle$. These non-Gaussian states are of metrological interest~\cite{Caves1981, PhysRevA.101.052331}, and their support on Fock states with $4k$ and $4k+2$ photons makes single-photon losses detectable by a parity measurement. The encoding requires $r\gtrsim2$ to reach fidelities above $0.99$ for arbitrary initial qubit states. We numerically confirmed that, under unitary evolution, the refocused conditional-squeezing gate reproduces the encoding fidelity of the ideal gate up to negligible deviations. This represents a substantial improvement over the scheme proposed in Ref.~\cite{PhysRevA.111.042606} that uses a controlled-squeezing gate, where the refocusing can not be applied because the time-reversal symmetry of the Hamiltonian is absent.

Including dissipation and the Kerr nonlinearity of the SQUID, with optimistic coherence times and Kerr strengths, the latter attainable in SNAIL-terminated resonators~\cite{Frattini2018, Sivak2019, Frattini2024, Lu2023}, we obtained encoding fidelities above $0.99$ for $\chi/2\pi\gtrsim12$~MHz, with photon loss in the resonator as the dominant error source. Compared with the algorithm of Ref.~\cite{PhysRevA.111.042606}, based on two applications of the controlled-squeezing gate, our protocol is twice as fast and achieves higher fidelities, significantly improving the performance of the encoding algorithm. As discussed in Appendix~\ref{apB}, the self-Kerr nonlinearity induced on the resonator by the dispersive coupling to the transmon, not included in our simulations, constitutes the main limitation of the present implementation for large squeezing parameters. 

We also extended the gate and the encoding algorithm to two resonator modes, where refocusing is again required. In this case, the logical states are even and odd superpositions of two-mode squeezed states, $|\chi_\pm\rangle_2$, and single-photon losses can be detected by measuring the photon-number difference between the modes. Due to the strong correlations of two-mode squeezed states, the two-mode gate reaches higher encoding fidelities with shorter gates, above $0.997$ for arbitrary qubit states with optimistic circuit parameters.


The refocusing technique presented here for the implementation of the conditional-squeezing gate has a broad range of potential applications. In particular, the refocusing approach could be generalized to prepare $D$ squeezed states along equally spaced directions in phase space, conditioned on the state of a qudit. The superpositions of these states, recently studied in Ref.~\cite{4d9d-kthj} and known as squeezed multiplets, belong to the class of rotation-symmetric bosonic codes~\cite{PhysRevX.10.011058}. This extension requires a D-tone flux drive and a corresponding time-inversion scheme, and is currently under study.

\hfill
\begin{acknowledgments}
This work was supported by Consejo Nacional de Investigaciones
Científicas y Técnicas (CONICET) and Universidad de
Buenos Aires (UBA).
\end{acknowledgments}

\appendix
\section{Hamiltonian of the SQUID-terminated resonator}
\label{apB}

In this Appendix we derive the Hamiltonian of Eq.~\eqref{full_hamiltonian} for the SQUID-terminated resonator of Figure~\ref{circuit}, following the approach of Ref.~\cite{PhysRevA.111.042606}. We start from the classical Lagrangian of the composite system, obtain the Hamiltonian for the resonator modes, and then quantize. We set $\hbar=1$. To avoid confusion with the external flux $\Phi(t)$, we denote the dimensionless phase field of the resonator by $\Theta(x,t)$.

\subsection{Lagrangian and mode equation}

The resonator is a transmission line of length $d$, with inductance $L_0$ and capacitance $C_0$ per unit length, terminated at $x=d$ by a SQUID with two identical junctions, each with Josephson energy $E_J$ and capacitance $C_J$. The Lagrangian is the sum of the transmission line and SQUID contributions (see, e.g., Ref.~\cite{PhysRevB.87.184501}),
\begin{align}
\mathcal{L} &= \left(\frac{1}{2e}\right)^{2}\frac{C_0}{2}\int_0^d \left(\dot\Theta^2 - v^2\,\Theta'^2\right)dx
+ \left(\frac{1}{2e}\right)^{2}\frac{2C_J}{2}\,\dot\Theta_d^{2} \nonumber\\
&\quad + 2E_J\cos\!\big(2e\,\Phi(t)\big)\cos\Theta_d ,
\label{app_lagrangian}
\end{align}
where $v=1/\sqrt{L_0C_0}$ is the propagation velocity, $\Theta_d(t)=\Theta(d,t)$, and dots and primes denote time and spatial derivatives, respectively. Here $\Phi(t)$ is the external flux through the SQUID, normalized such that the flux-dependent factor of the SQUID energy is $\cos(2e\Phi)$. The Euler-Lagrange equations in $x<d$ give the wave equation for the phase field
\begin{equation}
\ddot\Theta - v^2\Theta'' = 0,
\label{app_wave}
\end{equation}
\noindent together with the Neumann boundary condition at $x=0$, $\Theta'(0,t)=0$. The equation for $\Theta_d(t)$, on the other hand, requires a careful integration in the spatial coordinate of the Euler-Lagrange equation across $x=d$. This yields the following equation for the phase field at the SQUID position
\begin{equation}
\left(\frac{1}{2e}\right)^{2}\left[2C_J\,\ddot\Theta_d + C_0v^2\,\Theta'_d\right] + 2E_J\cos\!\big(2e\Phi(t)\big)\sin\Theta_d = 0 .
\label{app_boundary}
\end{equation}

For a static flux $\Phi(t)=\bar\Phi$ and $\Theta \ll 1$ (nonlinear corrections are discussed below), Eqs.~\eqref{app_wave} and \eqref{app_boundary} are solved by the following basis of solutions
\begin{equation}
\Theta(x,t)=e^{i\omega_n t}\psi_n(x),\qquad \psi_n(x)=\frac{1}{N_n}\cos(k_n x),
\label{app_modes}
\end{equation}
\noindent where $N_n$ is a normalization constant and $\omega_n=v k_n$ is the angular frequency for the modes. The wave numbers $k_n$ are the solutions of the transcendental equation determined by Eq.~\eqref{app_boundary}
\begin{equation}
\left(\frac{1}{2e}\right)^{2} C_0 v^2\, k\tan(kd) = 2E_J\cos(2e\bar\Phi) - \left(\frac{1}{2e}\right)^{2} 2C_J v^2 k^2 .
\label{app_transcendental}
\end{equation}

Eq.~\eqref{app_transcendental} determines the spectrum $\omega_n(\bar\Phi)$ for a given flux. It is important to stress that the above spatial modes are orthonormal with respect to the following inner product
\begin{equation}
\langle \psi_i,\psi_j\rangle=\int_0^d \epsilon(x)\,\psi_i(x)\psi_j(x)\,dx,
\label{app_inner}
\end{equation}
\noindent where $\epsilon(x)=\left(\frac{1}{2e}\right)^{2}\left[C_0+2C_J\,\delta(x-d)\right]$. This fixes $N_n^2=(2e)^{-2}\left\{\tfrac{C_0 d}{2}\left[1+\tfrac{\sin(2k_nd)}{2k_nd}\right]+2C_J\cos^2(k_nd)\right\}$.

\subsection{Time-dependent flux}

We now let $\Phi(t)=\Phi_s+\delta\Phi(t)$ with $|\delta\Phi|\ll \Phi_s$, so that $k_n$, and hence $\omega_n$ and $\psi_n$, become functions of time through Eq.~\eqref{app_transcendental}. These instantaneous modes remain orthonormal under Eq.~\eqref{app_inner} (the weight $\epsilon(x)$ does not depend on the flux), and we expand
\begin{equation}
\Theta(x,t)=\sum_n Q_n(t)\,\psi_n(t,x),
\label{app_expansion}
\end{equation}
\noindent where $Q_n(t)$ are time-dependent amplitudes. Using $\dot\psi_k=(\partial\psi_k/\partial\Phi)\,\delta\dot\Phi$, the kinetic term yields, together with the wave equation and the boundary condition (which cancel all terms localized at $x=d$), the Lagrangian
\begin{align}
\mathcal{L}=&\frac12\sum_i\left(\dot Q_i^2-\omega_i^2(\Phi)\,Q_i^2\right)+\delta\dot\Phi\sum_{ik}M_{ik}\,\dot Q_i Q_k \nonumber \\
& +\frac12\,\delta\dot\Phi^2\sum_{ijk} M_{ji}M_{jk}\,Q_iQ_k ,
\label{app_L_modes}
\end{align}
\noindent with the coupling matrix
\begin{equation}
M_{ik}=\Big\langle\psi_i,\frac{\partial\psi_k}{\partial\Phi}\Big\rangle .
\label{app_M}
\end{equation}

It is easy to see that the matrix $M_{ij}$ is antisymmetric, so $M_{ii}=0$. The canonical momenta are $P_i=\dot Q_i+\delta\dot\Phi\sum_k M_{ik}Q_k$, and the Legendre transform gives, to first order in $\delta\Phi$ and $\delta\dot\Phi$, the Hamiltonian
\begin{equation}
H=\frac12\sum_i\left(P_i^2+\omega_i^2(\Phi)\,Q_i^2\right)-\delta\dot\Phi\sum_{ik}M_{ik}\,P_iQ_k .
\label{app_H_classical}
\end{equation}

\subsection{Quantization and single-mode approximation}

We quantize the theory by promoting the generalized coordinates and momenta $Q_i$ and $P_i$ to operators $\hat Q_i$ and $\hat P_i$ acting on Hilbert space with canonical commutation rules, $[Q_i,P_j]=i\delta_{ij}$, and define the field operators of the \emph{static} resonator (i.e., when $\Phi=\Phi_s$),
\begin{equation}
\hat Q_i=\frac{\hat a_i+\hat a_i^\dagger}{\sqrt{2\bar\omega_i}},\qquad \hat P_i=-i\sqrt{\frac{\bar\omega_i}{2}}\,(\hat a_i-\hat a_i^\dagger),
\label{app_ladder}
\end{equation}
\noindent where $\bar\omega_i=\omega_i(\Phi_s)$. Expanding $\omega_i^2(\Phi)=\bar\omega_i^2+2\bar\omega_i\,\omega_i'\,\delta\Phi+O(\delta\Phi^2)$, with $\omega_i'=\partial\omega_i/\partial\Phi\,|_{\Phi_s}$, the Hamiltonian of Eq.~\eqref{app_H_classical} becomes
\begin{align}
\hat H&=\sum_i\left[\bar\omega_i\,\hat a_i^\dagger \hat a_i+\frac{\omega_i'}{2}\,\delta\Phi\,(\hat a_i+\hat a_i^\dagger)^2\right] \nonumber\\
&\quad-\delta\dot\Phi\sum_{i\neq k}\frac{M_{ik}}{2i}\sqrt{\frac{\bar\omega_i}{\bar\omega_k}}\left(\hat a_i\hat a_k-\hat a_i^\dagger \hat a_k+\hat a_i\hat a_k^\dagger-\hat a_i^\dagger \hat a_k^\dagger\right),
\label{app_H_quantum}
\end{align}
\noindent where we have dropped the constants. The first line is a set of harmonic oscillators whose frequencies are modulated by the flux; the second line couples different modes through $\delta\dot\Phi$, with coupling strengths determined by $M_{ik}$.

Henceforth, we assume that the drive is resonant with a single mode and all the other modes are far detuned from it. In this single-mode approximation, we obtain the quantum Hamiltonian
\begin{equation}
\hat H=\bar\omega\,\hat a^\dagger \hat a+\frac{\bar g_d}{2}\,\delta\Phi(t)\,(\hat a+\hat a^\dagger)^2,
\label{app_H_single}
\end{equation}
\noindent with $\bar g_d=(\partial \omega/\partial \Phi)|_{\Phi_s}$. This is the SQUID-terminated resonator part of Eq.~\eqref{full_hamiltonian}. Differentiating Eq.~\eqref{app_transcendental} gives explicitly
\begin{equation}
\bar g_d=-\,\frac{4e\,E_J\sin(2e\Phi_s)\,v}{\left(\frac{1}{2e}\right)^{2}\!\left[C_0v^2\left(\tan k d+\frac{kd}{\cos^2 kd}\right)+4C_Jv^2k\right]}\Bigg|_{k=\bar k},
\label{app_gd}
\end{equation}
so that $\bar g_d\propto\sin(2e\Phi_s)$ vanishes at the zero-flux sweet spot $\Phi_s=0$, and the SQUID must be biased at $\Phi_s\neq0$ for the squeezing drive to couple to the resonator.

\subsection{Nonlinear corrections and coupling to the qubit}

We have worked so far in the linear regime, $\Theta\ll1$. The leading correction from the expansion of $\cos\Theta_d$ is a SQUID-induced Kerr term $(K_\mathrm{SQUID}/2)\,\hat a^{\dagger2}\hat a^2$. It is controlled by the ratio $\sigma=E_{L}/(2E_J\cos(2e\Phi_s))$, where $E_L=(2e)^{-2}/(L_0d)$ is the inductive energy of the resonator, and, as shown in Ref.~\cite{PhysRevA.111.042606}, $K_\mathrm{SQUID}/\bar\omega\approx\sigma^{3}\pi Z_0/(2R_q)$, with $Z_0=\sqrt{L_0/C_0}$ and $R_q=1/(2e^2)$ the resistance quantum. For $\sigma\sim10^{-2}$ and $Z_0=50~\Omega$, this gives $K_\mathrm{SQUID}/2\pi$ of order $10^2$~Hz for $\bar\omega/2\pi=6$~GHz. We omitted this term in Eq.~\eqref{full_hamiltonian} and included it explicitly as $\hat H_\mathrm{Kerr}=(K/2)\,\hat a^{\dagger2}\hat a^2$ in the simulations of Sec.~\ref{open_system}.

The transmon is a weakly anharmonic oscillator with charging energy $E_C$, capacitively coupled to the resonator with coupling constant $g_q$ and detuning $\Delta_q=\omega_q-\omega$. In the dispersive regime, $g_q\ll|\Delta_q|$, the coupling can be eliminated to second order in $g_q/\Delta_q$, which yields the terms $\tfrac{\omega_q}{2}\hat\sigma_z+\tfrac{\chi}{2}\hat\sigma_z\hat a^\dagger\hat a$ of Eq.~\eqref{full_hamiltonian}, with

\begin{equation}
|\chi|=\frac{2g_q^2E_C}{|\Delta_q(\Delta_q-E_C)|} .
\label{app_chi}
\end{equation}

This approximation holds as long as the photon number in the resonator is well below the critical value $n_\mathrm{crit}=\Delta_q^2/(4g_q^2)$~\cite{Blais2021}. For $|\Delta_q|\gg E_C$, Eq.~\eqref{app_chi} gives $n_\mathrm{crit}\simeq E_C/(2|\chi|)$, independently of the detuning. Since a squeezed vacuum state contains $\langle\hat n\rangle=\sinh^2 r$ photons, the validity of Eq.~\eqref{full_hamiltonian} requires $|\chi|\sinh^2 r\ll E_C/2$. For a given transmon, this condition favors smaller dispersive shifts, at the cost of longer refocusing periods $T=\pi/|\chi|$.

Beyond second order in $g_q/\Delta_q$, the coupling to the transmon also induces a self-Kerr nonlinearity of the resonator, $K_q\simeq-\chi^2/(4E_C)$ for $|\Delta_q|\gg E_C$, whose value depends on the state of the qubit; equivalently, the dispersive shift itself becomes photon-number dependent. For typical parameters, $K_q$ is much larger than the SQUID-induced Kerr nonlinearity considered in Sec.~\ref{open_system}. Following the estimate of Sec.~\ref{open_system}, this nonlinearity tilts the squeezing direction by an angle $\delta\sim|K_q|\langle\hat n\rangle/g_d$, which must remain small compared with the angular width of the squeezed state, of order $1/\sinh(2r)$. Since $K_q\propto\chi^2$ while, for a fixed squeezing parameter and a fixed number of periods, $g_d\propto\chi$, reducing the dispersive shift relaxes this condition only linearly in $\chi/E_C$. This constitutes the main limitation of the present implementation for large squeezing parameters. It could be mitigated by operating at lower squeezing, at the cost of lower encoding fidelities.

The flux drive acts only on the resonator, so it does not directly excite the transmon. The higher transmon levels can instead become populated by the qubit operations of the protocol. Since the qubit frequency depends on the photon number through the dispersive shift, a $\pi$ pulse acting uniformly on all the photon-number components of the squeezed state must have a spectral width larger than $|\chi|\,\Delta n$, where $\Delta n=\sinh(2r)/\sqrt{2}$ is the width of the photon-number distribution of a squeezed vacuum state, whereas avoiding leakage to the second excited level requires, for simple pulse shapes, a spectral width small compared with $E_C$. This constraint is less demanding for the intermediate $\pi$ pulse, applied at $t=\tau$ when the squeezing parameter is half of its final value, than for the final one. Throughout this work, the $\pi$ pulses and Hadamard gates are assumed to be ideal.

\section{Two-mode SQUID-terminated resonator}
\label{apA}

Here, we consider the case of two relevant modes stored in the SQUID-terminated resonator, to which Sec.~\ref{tms_section} is devoted. In particular, we derive the interaction term between the modes, denoted $\hat H_{\text{int}}(t)$ in Eq.~\eqref{interaction_hamiltonian_tms}. From Eq.~\eqref{app_H_quantum}, and using that $M_{21}=-M_{12}$, the interaction Hamiltonian for two resonator modes reduces to

\begin{align}
\hat H_{\text{int}}(t)={}&-\delta\dot\Phi(t)\,\frac{M_{12}}{2i}\left[\left(\sqrt{\frac{\omega_1}{\omega_2}}-\sqrt{\frac{\omega_2}{\omega_1}}\right)\big(\hat a_1\hat a_2-\hat a^\dagger_1\hat a^\dagger_2\big)\right.\nonumber\\
&\left.+\left(\sqrt{\frac{\omega_1}{\omega_2}}+\sqrt{\frac{\omega_2}{\omega_1}}\right)\big(\hat a_1\hat a^\dagger_2-\hat a^\dagger_1\hat a_2\big)\right].
\end{align}

Defining the coupling constant between the modes as $\Gamma=M_{12}/(2\sqrt{\omega_1\omega_2})$, straightforward algebra yields the Hamiltonian of Eq.~\eqref{interaction_hamiltonian_tms}.

Considering a bichromatic driving of the form $\delta \Phi(t)=\epsilon_0\sin{(\omega_{d,0}t-\varphi_0)}+ \epsilon_1\sin{(\omega_{d,1}t-\varphi_1)}$, parametric resonance is driven with the choice $\omega_{d,0}=\omega_{1,+}+\omega_{2,+}$ and $\omega_{d,1}=\omega_{1,-}+\omega_{2,-}$ (see Sec.~\ref{subsection_tms}). In this picture, the bosonic operators of each mode become $\hat a_ie^{-i\omega_{i,\pm}t}$, where the $+$ ($-$) sign corresponds to the qubit state $\ket{0}$ ($\ket{1}$). For both qubit states, the two-mode squeezing term $\hat a_1\hat a_2-\hat a^\dagger_1\hat a^\dagger_2$ contributes with a static term (resonant to one of the driving frequencies) and a slowly-varying time-dependent contribution, which oscillate at frequency $\pm(\chi_1+\chi_2)$. The terms corresponding to the operator $\hat a_1\hat a^\dagger_2-\hat a^\dagger_1\hat a_2$ are fast oscillating and can be neglected within the RWA. This approximation leads to the Hamiltonian in Eq.~\eqref{approximate_hamiltonian_tms}.

We consider the following circuit parameters, which are well within reach of current experimental capabilities~\cite{Wulschner2016, Simoen2015}: $d=13.68$~mm, $Z_0=50~\Omega$, $v=1.2\times10^8$~m/s, $C_J=38.2$~fF, $E_J/h=0.1554$~THz, and $2e\Phi_s=0.2$~rad. With this geometry, we obtain mode frequencies $\omega_1/2\pi=6$~GHz and $\omega_2/2\pi=10$~GHz, and the interaction constant $\gamma=2\pi\times1.19$~MHz is reached with a drive amplitude $2e\epsilon_1\approx0.019$~rad, well within the weak-drive regime.

The Kerr nonlinearities induced in the resonator by the Josephson junctions of the SQUID can be obtained with the black-box quantization approach of Ref.~\cite{Nigg2012}, by expanding the Josephson potential $-2E_J\cos(2e\Phi_s)\cos\Theta_d$ to fourth order in the phase $\Theta_d$ across the SQUID. Each mode $n$ contributes to this phase according to its zero-point fluctuation $\theta_{\text{zpf},n}$, weighted by the mode function at the SQUID position,
\begin{equation}
\theta_{\text{zpf},n}\propto\lambda_n=\frac{\cos(k_nd)}{N_n\sqrt{\omega_n}}.
\end{equation}
Since the nonlinearity originates from a single lumped element coupled to both resonator modes, the self-Kerr strengths $K_n$ and the cross-Kerr coefficient are not independent. Expanding $\Theta_d^4=\big[\sum_n\theta_{\text{zpf},n}(\hat a_n+\hat a_n^\dagger)\big]^4$ and retaining only the terms that conserve the total photon number, one finds $K_n\propto\lambda_n^4$ and, for $n=1,2$,
\begin{equation}
\frac{K_1}{K_2}=\left(\frac{\lambda_1}{\lambda_2}\right)^4,
\end{equation}
which is entirely set by the geometry of the circuit. This relation, characteristic of a single Josephson element participating in a multimode circuit~\cite{Nigg2012}, fixes the two-mode Kerr spectrum once $K_1$ is specified. Finally, the cross-Kerr strength is
\begin{equation}
K_{12}=2\sqrt{K_1K_2}.
\end{equation}

\section{Master equation in the qubit-block representation}
\label{app:blockME}

In this Appendix we describe the method used to solve the master equation of Eq.~\eqref{master_equation}. We write the density matrix of the qubit--resonator system as

\begin{equation}
\hat\rho=|0\rangle\langle0|\otimes\hat\rho_{00}+|1\rangle\langle1|\otimes\hat\rho_{11}+|0\rangle\langle1|\otimes\hat\rho_{01}+|1\rangle\langle0|\otimes\hat\rho_{01}^\dagger,
\label{block_decomposition}
\end{equation}

\noindent where $\hat\rho_{jk}=\langle j|\hat\rho|k\rangle$ ($j,k=0,1$) are operators on the resonator Hilbert space. The diagonal blocks $\hat\rho_{00}$ and $\hat\rho_{11}$ are the (unnormalized) resonator states conditioned on each qubit state, whereas $\hat\rho_{01}$ contains the coherence between the two branches. Within the RWA, the Hamiltonian is diagonal in the qubit basis, $\hat H_I(t)+\hat H_{\text{Kerr}}=\sum_j|j\rangle\langle j|\otimes\hat H_{I,j}(t)$, where $\hat H_{I,j}(t)$ is the branch Hamiltonian that includes the Kerr term $(K/2)\hat n(\hat n-1)$. In this representation, photon loss and thermal excitation of the resonator act within each block, whereas the qubit channels couple them: pure dephasing damps the coherence $\hat\rho_{01}$, while relaxation and thermal excitation of the qubit transfer population between $\hat\rho_{11}$ and $\hat\rho_{00}$ and also damp $\hat\rho_{01}$. Explicitly, the master equation becomes

\begin{align}
\dot{\hat\rho}_{jk}={}&-i\big(\hat H_{I,j}\hat\rho_{jk}-\hat\rho_{jk}\hat H_{I,k}\big) \nonumber 
\\ 
& +\kappa_1\Big(e^{-i(\bar\omega_j-\bar\omega_k)t}\,\hat a\hat\rho_{jk}\hat a^\dagger-\tfrac12\{\hat a^{\dagger}\hat a,\hat\rho_{jk}\}\Big)\nonumber\\
&+\kappa_2\Big(e^{i(\bar\omega_j-\bar\omega_k)t}\,\hat a^\dagger\hat\rho_{jk}\hat a-\tfrac12\{\hat a\hat a^\dagger,\hat\rho_{jk}\}\Big)+\mathcal Q_{jk},
\label{block_ME}
\end{align}

\noindent where the phase factors arise from the time dependence of the jump operators $\hat a$ and $\hat a^\dagger$ in the interaction picture, and the qubit channels contribute

\begin{align}
\mathcal{Q}_{00}&=\kappa_4\,\hat U(t)\hat\rho_{11}\hat U^\dagger(t)-\kappa_3\hat\rho_{00},\\
\mathcal{Q}_{11}&=\kappa_3\,\hat U^\dagger(t)\hat\rho_{00}\hat U(t)-\kappa_4\hat\rho_{11},\\
\mathcal Q_{01}&=-\Big[\tfrac12(\kappa_3+\kappa_4)+2\kappa_5\Big]\hat\rho_{01},
\end{align}

\noindent with $\hat U(t)=e^{i\chi\hat n t}$.

The protocol is implemented as follows. The initial state $\hat H(\alpha|0\rangle+\beta|1\rangle)\otimes|0\rangle$, where $\hat H$ is the Hadamard gate, corresponds to $\hat\rho_{00}=|g|^2|0\rangle\langle0|$, $\hat\rho_{11}=|e|^2|0\rangle\langle0|$ and $\hat\rho_{01}=g e^*|0\rangle\langle0|$, with $g=(\alpha+\beta)/\sqrt2$ and $e=(\alpha-\beta)/\sqrt2$. Each $\pi$ pulse on the qubit maps $\hat\rho_{00}\leftrightarrow\hat\rho_{11}$ and $\hat\rho_{01}\to\hat\rho_{01}^\dagger$, and the drive phases are exchanged in the second half of the protocol. The final Hadamard gate yields the conditional resonator states

\begin{align}
\hat\rho'_{00}&=\tfrac12\big(\hat\rho_{00}+\hat\rho_{01}+\hat\rho_{01}^\dagger+\hat\rho_{11}\big),\\
\hat\rho'_{11}&=\tfrac12\big(\hat\rho_{00}-\hat\rho_{01}-\hat\rho_{01}^\dagger+\hat\rho_{11}\big),
\end{align}

whose traces are the probabilities of the two outcomes of the qubit measurement. The fidelity is then computed as $F=\langle\psi_0|\hat\rho'_{00}|\psi_0\rangle+\langle\psi_1|\hat\rho'_{11}|\psi_1\rangle$, with $|\psi_{0,1}\rangle$ the target encoded states for each measurement outcome ($|\psi_R\rangle$). This expression coincides with the average fidelity $\bar F=P_+F_++P_-F_-$ defined in Sec.~\ref{encoding_section}, averaged over all quantum jump evolutions.

Since $\hat H_{I,j}$ contains only the operators $\hat a^2$, $\hat a^{\dagger2}$ and $\hat n$, and the jump operators $\hat a$ and $\hat a^\dagger$ change the photon number by one on both sides of $\hat\rho_{jk}$, the matrix elements $\langle n|\hat\rho_{jk}|m\rangle$ with $n+m$ odd remain zero for an initial vacuum state, and only half of each block needs to be propagated. The equations were integrated with an adaptive eighth-order Runge--Kutta method, using a truncation of the resonator Hilbert space at $N=500$ for which the results are converged to $10^{-6}$. In the absence of dissipation, this method reproduces the results obtained by solving the Schrödinger equation with the full Hamiltonian of Eq.~\eqref{interaction_picture} within $10^{-4}$.

\nocite{*}

\bibliography{apssamp}

@PREAMBLE{
 "\providecommand{\noopsort}[1]{}" 
 # "\providecommand{\singleletter}[1]{#1}%" 
}

@article{Jiang2025,
  title = {Advancements in superconducting quantum computing},
  author = {Jiang, Yao-Yao and Deng, Chunqing and Fan, Heng and Li, Bing-Yang and Sun, Luyan and Tan, Xin-Sheng and Wang, Weiting and Xue, Guang-Ming and Yan, Fei and Yu, Hai-Feng and Zhang, Ying-Shan and Zhang, Yu-Ran and Zou, Chang-Ling},
  journal = {National Science Review},
  volume = {12},
  number = {8},
  pages = {nwaf246},
  year = {2025},
  doi = {10.1093/nsr/nwaf246}
}

@article{Ourjoumtsev2006,
  title = {Generating Optical Schr{\"o}dinger Kittens for Quantum Information Processing},
  author = {Ourjoumtsev, Alexei and Tualle-Brouri, Rosa and Laurat, Julien and Grangier, Philippe},
  journal = {Science},
  volume = {312},
  pages = {83--86},
  year = {2006},
  doi = {10.1126/science.1122858}
}

@article{Knill2001,
  title = {A scheme for efficient quantum computation with linear optics},
  author = {Knill, E. and Laflamme, R. and Milburn, G. J.},
  journal = {Nature},
  volume = {409},
  pages = {46--52},
  year = {2001},
  doi = {10.1038/35051009}
}

@article{Pittman2005,
  title = {Experimental demonstration of a quantum circuit using linear optics gates},
  author = {Pittman, T. B. and Jacobs, B. C. and Franson, J. D.},
  journal = {Phys. Rev. A},
  volume = {71},
  pages = {032307},
  year = {2005},
  doi = {10.1103/PhysRevA.71.032307}
}

@article{Leindecker2026,
  title = {State-dependent Gaussian gate set using an optical tweezer for trapped ions},
  author = {Leindecker, Philip and Milanovic, Luka and Behrle, Tanja and Brucke, Edgar and Marinelli, Matteo and Schmidt, Julian and Home, Jonathan and Hempel, Cornelius},
  journal = {arXiv:2606.31864},
  year = {2026}
}

@article{CiracZoller1995,
  title = {Quantum Computations with Cold Trapped Ions},
  author = {Cirac, J. I. and Zoller, P.},
  journal = {Phys. Rev. Lett.},
  volume = {74},
  pages = {4091--4094},
  year = {1995},
  doi = {10.1103/PhysRevLett.74.4091}
}

@article{Jaksch2000,
  title = {Fast Quantum Gates for Neutral Atoms},
  author = {Jaksch, D. and Cirac, J. I. and Zoller, P. and Rolston, S. L. and C{\^o}t{\'e}, R. and Lukin, M. D.},
  journal = {Phys. Rev. Lett.},
  volume = {85},
  pages = {2208--2211},
  year = {2000},
  doi = {10.1103/PhysRevLett.85.2208}
}

@article{Isenhower2010,
  title = {Demonstration of a Neutral Atom Controlled-NOT Quantum Gate},
  author = {Isenhower, L. and Urban, E. and Zhang, X. L. and Gill, A. T. and Henage, T. and Johnson, T. A. and Walker, T. G. and Saffman, M.},
  journal = {Phys. Rev. Lett.},
  volume = {104},
  pages = {010503},
  year = {2010},
  doi = {10.1103/PhysRevLett.104.010503}
}

@article{Evered2023,
  title = {High-fidelity parallel entangling gates on a neutral-atom quantum computer},
  author = {Evered, S. J. and Bluvstein, D. and Kalinowski, M. and Ebadi, S. and Manovitz, T. and Zhou, H. and others},
  journal = {Nature},
  volume = {622},
  pages = {268--272},
  year = {2023},
  doi = {10.1038/s41586-023-06481-y}
}

@article{Devoret2013,
  title = {Superconducting Circuits for Quantum Information: An Outlook},
  author = {Devoret, M. H. and Schoelkopf, R. J.},
  journal = {Science},
  volume = {339},
  pages = {1169--1174},
  year = {2013},
  doi = {10.1126/science.1231930}
}

@article{PhysRevA.101.052331,
  title = {State-dependent motional squeezing of a trapped ion: Proposed method and applications},
  author = {Drechsler, Mart\'{\i}n and Bel\'en Far\'{\i}as, M. and Freitas, Nahuel and Schmiegelow, Christian T. and Paz, Juan Pablo},
  journal = {Phys. Rev. A},
  volume = {101},
  issue = {5},
  pages = {052331},
  numpages = {4},
  year = {2020},
  month = {May},
  publisher = {American Physical Society},
  doi = {10.1103/PhysRevA.101.052331},
  url = {https://link.aps.org/doi/10.1103/PhysRevA.101.052331}
}

@article{Madsen2022,
  title = {Quantum computational advantage with a programmable photonic processor},
  author = {Madsen, Lars S. and others},
  journal = {Nature},
  volume = {606},
  pages = {75--81},
  year = {2022},
  doi = {10.1038/s41586-022-04725-x}
}

@article{Google2024,
  title = {Quantum error correction below the surface code threshold},
  author = {{Google Quantum AI and Collaborators}},
  journal = {Nature},
  volume = {638},
  pages = {920--926},
  year = {2024},
  doi = {10.1038/s41586-024-08449-y}
}

@article{Google2023,
  title = {Suppressing quantum errors by scaling a surface code logical qubit},
  author = {{Google Quantum AI}},
  journal = {Nature},
  volume = {614},
  pages = {676--681},
  year = {2023},
  doi = {10.1038/s41586-022-05434-1}
}

@article{Bluvstein2024,
  title = {Logical quantum processor based on reconfigurable atom arrays},
  author = {Bluvstein, Dolev and Evered, Simon J. and Geim, Alexandra A. and Li, Sophie H. and Zhou, Hengyun and Manovitz, Tom and Ebadi, Sepehr and Cain, Madelyn and Kalinowski, Marcin and Hangleiter, Dominik and Bonilla Ataides, J. Pablo and Maskara, Nishad and Cong, Iris and Gao, Xun and Sales Rodriguez, Pedro and Karolyshyn, Thomas and Semeghini, Giulia and Gullans, Michael J. and Greiner, Markus and Vuletić, Vladan and Lukin, Mikhail D.},
  journal = {Nature},
  volume = {626},
  pages = {58--65},
  year = {2024},
  doi = {10.1038/s41586-023-06927-3}
}

@article{Butt2026,
  title = {Demonstration of measurement-free universal logical quantum computation},
  author = {Butt, Friederike and Pogorelov, Ivan and Freund, Robert and Steiner, Alex and Meyer, Marcel and Monz, Thomas and M{\"u}ller, Markus},
  journal = {Nature Communications},
  year = {2026},
  doi = {10.1038/s41467-026-68533-x}
}

@article{Zhang2025,
  title = {Demonstrating a universal logical gate set in error-detecting surface codes on a superconducting quantum processor},
  author = {Zhang, Jiaxuan and Chen, Zhao-Yun and Wang, Yun-Jie and Lu, Bin-Han and Zhang, Hai-Feng and Li, Jia-Ning and Duan, Peng and Wu, Yu-Chun and Guo, Guo-Ping},
  journal = {npj Quantum Information},
  volume = {11},
  pages = {177},
  year = {2025},
  doi = {10.1038/s41534-025-01118-6}
}

@article{Lamata2026,
  title = {Digital--Analog Quantum Simulation and Computing: A Perspective on Past and Future Developments},
  author = {Lamata, Lucas},
  journal = {Advanced Computing},
  volume = {1},
  year = {2026},
  doi = {10.1002/adco.70002}
}

@article{Kim2023,
  title = {Evidence for the utility of quantum computing before fault tolerance},
  author = {Kim, Youngseok and Eddins, Andrew and Anand, Sajant and Wei, Ken Xuan and van den Berg, Ewout and Rosenblatt, Sami and Nayfeh, Hasan and Wu, Yantao and Zaletel, Michael and Temme, Kristan and Kandala, Abhinav},
  journal = {Nature},
  volume = {618},
  pages = {500--505},
  year = {2023},
  doi = {10.1038/s41586-023-06096-3}
}

@article{PhysRevLett.82.1784,
  title = {Quantum Computation over Continuous Variables},
  author = {Lloyd, Seth and Braunstein, Samuel L.},
  journal = {Phys. Rev. Lett.},
  volume = {82},
  issue = {8},
  pages = {1784--1787},
  numpages = {0},
  year = {1999},
  month = {Feb},
  publisher = {American Physical Society},
  doi = {10.1103/PhysRevLett.82.1784},
  url = {https://link.aps.org/doi/10.1103/PhysRevLett.82.1784}
}

@article{Eriksson2024,
  title = {Universal control of a bosonic mode via drive-activated native cubic interactions},
  author = {Eriksson, Axel M. and S{\'e}pulcre, Th{\'e}o and Kervinen, Mikael and Hillmann, Timo and Kudra, Marina and Dupouy, Simon and Lu, Yong and Khanahmadi, Maryam and Yang, Jiaying and Castillo-Moreno, Claudia and Delsing, Per and Gasparinetti, Simone},
  journal = {Nature Communications},
  volume = {15},
  pages = {2512},
  year = {2024},
  doi = {10.1038/s41467-024-46507-1}
}

@article{Hillmann2020,
  title = {Universal Gate Set for Continuous-Variable Quantum Computation with Microwave Circuits},
  author = {Hillmann, Timo and Quijandr{\'i}a, Fernando and Johansson, G{\"o}ran and Ferraro, Alessandro and Gasparinetti, Simone and Ferrini, Giulia},
  journal = {Phys. Rev. Lett.},
  volume = {125},
  pages = {160501},
  year = {2020},
  doi = {10.1103/PhysRevLett.125.160501}
}

@article{Grimm2020,
  title = {Stabilization and operation of a Kerr-cat qubit},
  author = {Grimm, A. and Frattini, N. E. and Puri, S. and Mundhada, S. O. and Devoret, M. H. and others},
  journal = {Nature},
  volume = {584},
  pages = {205--209},
  year = {2020},
  doi = {10.1038/s41586-020-2587-z}
}

@article{Heeres2015,
  title = {Cavity State Manipulation Using Photon-Number Selective Phase Gates},
  author = {Heeres, Reinier W. and Vlastakis, Brian and Holland, Eric and Krastanov, Stefan and Albert, Victor V. and Frunzio, Luigi and Jiang, Liang and Schoelkopf, Robert J.},
  journal = {Phys. Rev. Lett.},
  volume = {115},
  pages = {137002},
  year = {2015},
  doi = {10.1103/PhysRevLett.115.137002}
}

@article{Kudra2022,
  title = {Robust Preparation of Wigner-Negative States with Optimized SNAP-Displacement Sequences},
  author = {Kudra, M. and Kervinen, M. and Strandberg, I. and Ahmed, S. and Scigliuzzo, M. and Osman, A. and Lozano, D. P. and Tholén, M. O. and Borgani, R. and Haviland, D. B. and Ferrini, G. and Bylander, J. and Kockum, A. F. and Quijandría, F. and Delsing, P. and Gasparinetti, S.},
  journal = {PRX Quantum},
  volume = {3},
  pages = {030301},
  year = {2022},
  doi = {10.1103/PRXQuantum.3.030301}
}

@ARTICLE{PhysRevA.111.042606,
  title = {Controlled-squeeze gate in superconducting quantum circuits},
  author = {Del Grosso, Nicol\'as F. and Corti\~nas, Rodrigo G. and Villar, Paula I. and Lombardo, Fernando C. and Paz, Juan Pablo},
  journal = {Phys. Rev. A},
  volume = {111},
  issue = {4},
  pages = {042606},
  numpages = {9},
  year = {2025},
  month = {Apr},
  publisher = {American Physical Society},
  doi = {10.1103/PhysRevA.111.042606},
  url = {https://link.aps.org/doi/10.1103/PhysRevA.111.042606}
}

@article{JJ,
  title = {Josephson-junction-embedded transmission-line resonators: From Kerr medium to in-line transmon},
  author = {Bourassa, J. and Beaudoin, F. and Gambetta, Jay M. and Blais, A.},
  journal = {Phys. Rev. A},
  volume = {86},
  issue = {1},
  pages = {013814},
  numpages = {13},
  year = {2012},
  month = {Jul},
  publisher = {American Physical Society},
  doi = {10.1103/PhysRevA.86.013814},
  url = {https://link.aps.org/doi/10.1103/PhysRevA.86.013814}
}

@article{Nigg,
  title = {Black-Box Superconducting Circuit Quantization},
  author = {Nigg, Simon E. and Paik, Hanhee and Vlastakis, Brian and Kirchmair, Gerhard and Shankar, S. and Frunzio, Luigi and Devoret, M. H. and Schoelkopf, R. J. and Girvin, S. M.},
  journal = {Phys. Rev. Lett.},
  volume = {108},
  issue = {24},
  pages = {240502},
  numpages = {5},
  year = {2012},
  month = {Jun},
  publisher = {American Physical Society},
  doi = {10.1103/PhysRevLett.108.240502},
  url = {https://link.aps.org/doi/10.1103/PhysRevLett.108.240502}
}

@article{PhysRevB.87.184501,
  title = {Parametric resonance in tunable superconducting cavities},
  author = {Wustmann, Waltraut and Shumeiko, Vitaly},
  journal = {Phys. Rev. B},
  volume = {87},
  issue = {18},
  pages = {184501},
  numpages = {23},
  year = {2013},
  month = {May},
  publisher = {American Physical Society},
  doi = {10.1103/PhysRevB.87.184501},
  url = {https://link.aps.org/doi/10.1103/PhysRevB.87.184501}
}

@article{Krantz2013,
  title = {Investigation of nonlinear effects in {J}osephson parametric oscillators used in circuit quantum electrodynamics},
  author = {Krantz, Philip and Reshitnyk, Yarema and Wustmann, Waltraut and Bylander, Jonas and Gustavsson, Simon and Oliver, William D. and Duty, Timothy and Shumeiko, Vitaly and Delsing, Per},
  journal = {New Journal of Physics},
  volume = {15},
  number = {10},
  pages = {105002},
  year = {2013},
  month = oct,
  publisher = {IOP Publishing},
  doi = {10.1088/1367-2630/15/10/105002},
  url = {https://doi.org/10.1088/1367-2630/15/10/105002}
}

@article{Frattini2018,
  title = {Optimizing the nonlinearity and dissipation of a {SNAIL} parametric amplifier for dynamic range},
  author = {Frattini, N. E. and Sivak, V. V. and Lingenfelter, A. and Shankar, S. and Devoret, M. H.},
  journal = {Physical Review Applied},
  volume = {10},
  number = {5},
  pages = {054020},
  year = {2018},
  month = nov,
  publisher = {American Physical Society},
  doi = {10.1103/PhysRevApplied.10.054020}
}

@article{Sivak2019,
  title = {Kerr-free three-wave mixing in superconducting quantum circuits},
  author = {Sivak, V. V. and Frattini, N. E. and Joshi, V. R. and Lingenfelter, A. and Shankar, S. and Devoret, M. H.},
  journal = {Physical Review Applied},
  volume = {11},
  number = {5},
  pages = {054060},
  year = {2019},
  month = may,
  publisher = {American Physical Society},
  doi = {10.1103/PhysRevApplied.11.054060}
}

@article{Frattini2024,
  title = {Observation of pairwise level degeneracies and the quantum regime of the {A}rrhenius law in a double-well parametric oscillator},
  author = {Frattini, N. E. and Corti{\~n}as, R. G. and Venkatraman, J. and Xiao, X. and Su, Q. and Lei, C. U. and Chapman, B. J. and Joshi, V. R. and Girvin, S. M. and Schoelkopf, R. J. and Puri, S. and Devoret, M. H.},
  journal = {Physical Review X},
  volume = {14},
  number = {3},
  pages = {031040},
  year = {2024},
  publisher = {American Physical Society},
  doi = {10.1103/PhysRevX.14.031040}
}

@article{Lu2023,
  title = {Resolving {F}ock states near the {K}err-free point of a superconducting resonator},
  author = {Lu, Yong and Kudra, Marina and Hillmann, Timo and Yang, Jiaying and Li, Hangxi and Quijandr{\'i}a, Fernando and Delsing, Per},
  journal = {npj Quantum Information},
  volume = {9},
  number = {1},
  pages = {114},
  year = {2023},
  publisher = {Nature Publishing Group},
  doi = {10.1038/s41534-023-00782-w}
}

@article{Blais2021,
  title = {Circuit quantum electrodynamics},
  author = {Blais, Alexandre and Grimsmo, Arne L. and Girvin, S. M. and Wallraff, Andreas},
  journal = {Rev. Mod. Phys.},
  volume = {93},
  issue = {2},
  pages = {025005},
  numpages = {72},
  year = {2021},
  month = {May},
  publisher = {American Physical Society},
  doi = {10.1103/RevModPhys.93.025005},
  url = {https://link.aps.org/doi/10.1103/RevModPhys.93.025005}
}

@article{PhysRevLett.130.267001,
  title = {Millisecond Coherence in a Superconducting Qubit},
  author = {Somoroff, Aaron and Ficheux, Quentin and Mencia, Raymond A. and Xiong, Haonan and Kuzmin, Roman and Manucharyan, Vladimir E.},
  journal = {Phys. Rev. Lett.},
  volume = {130},
  issue = {26},
  pages = {267001},
  numpages = {6},
  year = {2023},
  month = {Jun},
  publisher = {American Physical Society},
  doi = {10.1103/PhysRevLett.130.267001},
  url = {https://link.aps.org/doi/10.1103/PhysRevLett.130.267001}
}

@article{Bland2025,
  title = {Millisecond lifetimes and coherence times in {2D} transmon qubits},
  author = {Bland, Matthew P. and Bahrami, Faranak and Martinez, Jeronimo G. C. and Prestegaard, Paal H. and Smitham, Basil M. and Joshi, Atharv and Hedrick, Elizabeth and Kumar, Shashwat and Yang, Ambrose and Pakpour-Tabrizi, Alexander C. and Jindal, Apoorv and Chang, Ray D. and Cheng, Guangming and Yao, Nan and Cava, Robert J. and de Leon, Nathalie P. and Houck, Andrew A.},
  journal = {Nature},
  volume = {647},
  number = {8089},
  pages = {343--348},
  year = {2025},
  publisher = {Nature Publishing Group},
  doi = {10.1038/s41586-025-09687-4}
}

@article{Ganjam2024,
  author  = {Ganjam, Suhas and Wang, Yanhao and Lu, Yao and Banerjee, Archan and Lei, Chan U. and Krayzman, Lev and Kisslinger, Kim and Zhou, Chenyu and Li, Ruoshui and Jia, Yichen and Liu, Mingzhao and Frunzio, Luigi and Schoelkopf, Robert J.},
  title   = {Surpassing millisecond coherence in on-chip superconducting quantum memories by optimizing materials and circuit design},
  journal = {Nature Communications},
  volume  = {15},
  pages   = {3687},
  year    = {2024},
  doi     = {10.1038/s41467-024-47857-6}
}

@article{CavesSchumaker1985,
  author  = {Caves, Carlton M. and Schumaker, Bonny L.},
  title   = {New formalism for two-photon quantum optics. I. Quadrature phases and squeezed states},
  journal = {Physical Review A},
  volume  = {31},
  pages   = {3068--3092},
  year    = {1985},
  doi     = {10.1103/PhysRevA.31.3068}
}

@article{Weedbrook2012,
  author  = {Weedbrook, Christian and Pirandola, Stefano and Garc{\'i}a-Patr{\'o}n, Ra{\'u}l and Cerf, Nicolas J. and Ralph, Timothy C. and Shapiro, Jeffrey H. and Lloyd, Seth},
  title   = {Gaussian quantum information},
  journal = {Reviews of Modern Physics},
  volume  = {84},
  pages   = {621--669},
  year    = {2012},
  doi     = {10.1103/RevModPhys.84.621}
}

@article{Takeda2013,
  author  = {Takeda, Shuntaro and Mizuta, Takahiro and Fuwa, Maria and van Loock, Peter and Furusawa, Akira},
  title   = {Deterministic quantum teleportation of photonic quantum bits by a hybrid technique},
  journal = {Nature},
  volume  = {500},
  pages   = {315--318},
  year    = {2013},
  doi     = {10.1038/nature12366}
}

@article{Li2002,
  author  = {Li, Xiaoying and Pan, Qing and Jing, Jietai and Zhang, Jing and Xie, Changde and Peng, Kunchi},
  title   = {Quantum Dense Coding Exploiting a Bright Einstein-Podolsky-Rosen Beam},
  journal = {Physical Review Letters},
  volume  = {88},
  pages   = {047904},
  year    = {2002},
  doi     = {10.1103/PhysRevLett.88.047904}
}

@article{BraunsteinKimble2000,
  author  = {Braunstein, Samuel L. and Kimble, H. J.},
  title   = {Dense coding for continuous variables},
  journal = {Physical Review A},
  volume  = {61},
  pages   = {042302},
  year    = {2000},
  doi     = {10.1103/PhysRevA.61.042302}
}

@article{PhysRevLett.97.110501,
  title = {Universal Quantum Computation with Continuous-Variable Cluster States},
  author = {Menicucci, Nicolas C. and van Loock, Peter and Gu, Mile and Weedbrook, Christian and Ralph, Timothy C. and Nielsen, Michael A.},
  journal = {Phys. Rev. Lett.},
  volume = {97},
  issue = {11},
  pages = {110501},
  numpages = {4},
  year = {2006},
  month = {Sep},
  publisher = {American Physical Society},
  doi = {10.1103/PhysRevLett.97.110501},
  url = {https://link.aps.org/doi/10.1103/PhysRevLett.97.110501}
}

@article{Asavanant2019,
  author  = {Asavanant, Warit and Shiozawa, Yu and Yokoyama, Shota and Charoensombutamon, Baramee and Emura, Hiroki and Alexander, Rafael N. and Takeda, Shuntaro and Yoshikawa, Jun-ichi and Menicucci, Nicolas C. and Yonezawa, Hidehiro and Furusawa, Akira},
  title   = {Generation of time-domain-multiplexed two-dimensional cluster state},
  journal = {Science},
  volume  = {366},
  pages   = {373--376},
  year    = {2019},
  doi     = {10.1126/science.aay2645}
}

@article{Larsen2019,
  author  = {Larsen, Mikkel V. and Guo, Xueshi and Breum, Casper R. and Neergaard-Nielsen, Jonas S. and Andersen, Ulrik L.},
  title   = {Deterministic generation of a two-dimensional cluster state},
  journal = {Science},
  volume  = {366},
  pages   = {369--372},
  year    = {2019},
  doi     = {10.1126/science.aay4354}
}

@article{Wulschner2016,
  author  = {Wulschner, Friedrich and Goetz, Jan and Koessel, Fabian R. and others},
  title   = {Tunable coupling of transmission-line microwave resonators mediated by an rf SQUID},
  journal = {EPJ Quantum Technology},
  volume  = {3},
  pages   = {10},
  year    = {2016},
  doi     = {10.1140/epjqt/s40507-016-0048-2}
}

@article{Nigg2012,
  author  = {Nigg, Simon E. and Paik, Hanhee and Vlastakis, Brian and Kirchmair, Gerhard and Shankar, S. and Frunzio, Luigi and Devoret, Michel H. and Schoelkopf, Robert J. and Girvin, S. M.},
  title   = {Black-Box Superconducting Circuit Quantization},
  journal = {Physical Review Letters},
  volume  = {108},
  pages   = {240502},
  year    = {2012},
  doi     = {10.1103/PhysRevLett.108.240502}
}

@article{Caves1981,
  author  = {Caves, Carlton M.},
  title   = {Quantum-mechanical noise in an interferometer},
  journal = {Physical Review D},
  volume  = {23},
  pages   = {1693--1708},
  year    = {1981},
  doi     = {10.1103/PhysRevD.23.1693}
}

@article{Heeres2017,
  author  = {Heeres, Reinier W. and Reinhold, Philip and Ofek, Nissim and Frunzio, Luigi and Jiang, Liang and Devoret, Michel H. and Schoelkopf, Robert J.},
  title   = {Implementing a universal gate set on a logical qubit encoded in an oscillator},
  journal = {Nature Communications},
  volume  = {8},
  pages   = {94},
  year    = {2017},
  doi     = {10.1038/s41467-017-00045-1}
}

@article{Krastanov2015,
  author  = {Krastanov, Stefan and Albert, Victor V. and Shen, Chao and Zou, Chang-Ling and Heeres, Reinier W. and Vlastakis, Brian and Schoelkopf, Robert J. and Jiang, Liang},
  title   = {Universal control of an oscillator with dispersive coupling to a qubit},
  journal = {Physical Review A},
  volume  = {92},
  pages   = {040303},
  year    = {2015},
  doi     = {10.1103/PhysRevA.92.040303}
}

@article{Eickbusch2022,
  author  = {Eickbusch, Alec and Sivak, Volodymyr and Ding, Andy Z. and Elder, Salvatore S. and Jha, Shantanu R. and Venkatraman, Jayameenakshi and Royer, Baptiste and Girvin, S. M. and Schoelkopf, Robert J. and Devoret, Michel H.},
  title   = {Fast universal control of an oscillator with weak dispersive coupling to a qubit},
  journal = {Nature Physics},
  volume  = {18},
  pages   = {1464--1469},
  year    = {2022},
  doi     = {10.1038/s41567-022-01776-9}
}

@article{Hahn1950,
  author  = {Hahn, E. L.},
  title   = {Spin Echoes},
  journal = {Physical Review},
  volume  = {80},
  pages   = {580--594},
  year    = {1950},
  doi     = {10.1103/PhysRev.80.580}
}

@article{CarrPurcell1954,
  author  = {Carr, H. Y. and Purcell, E. M.},
  title   = {Effects of Diffusion on Free Precession in Nuclear Magnetic Resonance Experiments},
  journal = {Physical Review},
  volume  = {94},
  pages   = {630--638},
  year    = {1954},
  doi     = {10.1103/PhysRev.94.630}
}

@article{Sheldon2016,
  author  = {Sheldon, Sarah and Magesan, Easwar and Chow, Jerry M. and Gambetta, Jay M.},
  title   = {Procedure for systematically tuning up crosstalk in the cross-resonance gate},
  journal = {Physical Review A},
  volume  = {93},
  pages   = {060302},
  year    = {2016},
  doi     = {10.1103/PhysRevA.93.060302}
}

@article{Simoen2015,
  author = {Simoen, M. and Chang, C. W. S. and Krantz, P. and Bylander, J. and Wustmann, W. and Shumeiko, V. and Delsing, P. and Wilson, C. M.},
  title = {Characterization of a multimode coplanar waveguide parametric amplifier},
  journal = {J. Appl. Phys.},
  volume = {118},
  pages = {154501},
  year = {2015},
  doi = {10.1063/1.4933265}
}

@article{4d9d-kthj,
  title = {Squeezed quantum multiplets: Properties and phase space representation},
  author = {Paz, Juan Pablo and R{\'e}vora, Corina and Schmiegelow, Christian Tom{\'a}s},
  journal = {Phys. Rev. A},
  volume = {114},
  issue = {2},
  pages = {022459},
  numpages = {10},
  year = {2026},
  month = {Aug},
  publisher = {American Physical Society},
  doi = {10.1103/4d9d-kthj},
  url = {https://link.aps.org/doi/10.1103/4d9d-kthj}
}

@article{PhysRevX.10.011058,
  title = {Quantum Computing with Rotation-Symmetric Bosonic Codes},
  author = {Grimsmo, Arne L. and Combes, Joshua and Baragiola, Ben Q.},
  journal = {Phys. Rev. X},
  volume = {10},
  issue = {1},
  pages = {011058},
  numpages = {32},
  year = {2020},
  month = {Mar},
  publisher = {American Physical Society},
  doi = {10.1103/PhysRevX.10.011058},
  url = {https://link.aps.org/doi/10.1103/PhysRevX.10.011058}
}

@article{PhysRevLett.82.2417,
  title = {Dynamical Decoupling of Open Quantum Systems},
  author = {Viola, Lorenza and Knill, Emanuel and Lloyd, Seth},
  journal = {Phys. Rev. Lett.},
  volume = {82},
  issue = {12},
  pages = {2417--2421},
  numpages = {0},
  year = {1999},
  month = {Mar},
  publisher = {American Physical Society},
  doi = {10.1103/PhysRevLett.82.2417},
  url = {https://link.aps.org/doi/10.1103/PhysRevLett.82.2417}
}

@article{Joshi2021,
  author = {Joshi, Atharv and Noh, Kyungjoo and Gao, Yvonne Y.},
  title = {Quantum information processing with bosonic qubits in circuit QED},
  journal = {Quantum Sci. Technol.},
  volume = {6},
  number = {3},
  pages = {033001},
  year = {2021},
  doi = {10.1088/2058-9565/abe989}
}

@article{Wilson2011,
  author = {Wilson, C. M. and Johansson, G. and Pourkabirian, A. and Simoen, M. and Johansson, J. R. and Duty, T. and Nori, F. and Delsing, P.},
  title = {Observation of the dynamical Casimir effect in a superconducting circuit},
  journal = {Nature},
  volume = {479},
  number = {7373},
  pages = {376--379},
  year = {2011},
  doi = {10.1038/nature10561}
}

@article{4ts4-qj74,
  title = {Observation of Genuine Tripartite Non-Gaussian Entanglement from a Superconducting Three-Photon Spontaneous Parametric Down-Conversion Source},
  author = {Jarvis-Frain, Benjamin and Schang, Andy and Quijandr\'{\i}a, Fernando and Nsanzineza, Ibrahim and Dubyna, Dmytro and Chang, C. W. Sandbo and Nori, Franco and Wilson, C. M.},
  journal = {Phys. Rev. Lett.},
  volume = {137},
  issue = {4},
  pages = {040201},
  numpages = {7},
  year = {2026},
  month = {Jul},
  publisher = {American Physical Society},
  doi = {10.1103/4ts4-qj74},
  url = {https://link.aps.org/doi/10.1103/4ts4-qj74}
}

@article{Lienhard2025,
  author = {Lienhard, Vincent and Martin, Romain and Chew, Yuki Torii and Tomita, Takafumi and Ohmori, Kenji and de L{\'e}s{\'e}leuc, Sylvain},
  title = {Generation of Motional Squeezed States for Neutral Atoms in Optical Tweezers},
  journal = {Phys. Rev. Lett.},
  volume = {135},
  pages = {253404},
  year = {2025},
  doi = {10.1103/3kwz-ny2h}
}

@article{Kurpiers2018,
  author = {Kurpiers, P. and Magnard, P. and Walter, T. and Royer, B. and Pechal, M. and Heinsoo, J. and Salath{\'e}, Y. and Akin, A. and Storz, S. and Besse, J.-C. and Gasparinetti, S. and Blais, A. and Wallraff, A.},
  title = {Deterministic quantum state transfer and remote entanglement using microwave photons},
  journal = {Nature},
  volume = {558},
  pages = {264--267},
  year = {2018},
  doi = {10.1038/s41586-018-0195-y}
}

@article{Ofek2016,
  author = {Ofek, Nissim and Petrenko, Andrei and Heeres, Reinier and Reinhold, Philip and Leghtas, Zaki and Vlastakis, Brian and Liu, Yehan and Frunzio, Luigi and Girvin, S. M. and Jiang, L. and Mirrahimi, Mazyar and Devoret, M. H. and Schoelkopf, R. J.},
  title = {Extending the lifetime of a quantum bit with error correction in superconducting circuits},
  journal = {Nature},
  volume = {536},
  pages = {441--445},
  year = {2016},
  doi = {10.1038/nature18949}
}

@article{Burd2019,
  author = {Burd, S. C. and Srinivas, R. and Bollinger, J. J. and Wilson, A. C. and Wineland, D. J. and Leibfried, D. and Slichter, D. H. and Allcock, D. T. C.},
  title = {Quantum amplification of mechanical oscillator motion},
  journal = {Science},
  volume = {364},
  pages = {1163--1165},
  year = {2019},
  doi = {10.1126/science.aaw2884}
}

@article{Motzoi2009,
  title = {Simple Pulses for Elimination of Leakage in Weakly Nonlinear Qubits},
  author = {Motzoi, F. and Gambetta, J. M. and Rebentrost, P. and Wilhelm, F. K.},
  journal = {Phys. Rev. Lett.},
  volume = {103},
  pages = {110501},
  year = {2009},
  doi = {10.1103/PhysRevLett.103.110501}
}

@article{Ye2021,
  title = {Engineering Purely Nonlinear Coupling between Superconducting Qubits Using a Quarton},
  author = {Ye, Yufeng and Peng, Kaidong and Naghiloo, Mahdi and Cunningham, Gregory and O'Brien, Kevin P.},
  journal = {Phys. Rev. Lett.},
  volume = {127},
  pages = {050502},
  year = {2021},
  doi = {10.1103/PhysRevLett.127.050502}
}

@article{Ye2024,
  title = {Ultrafast superconducting qubit readout with the quarton coupler},
  author = {Ye, Yufeng and Kline, Jeremy B. and Chen, Sean and Yen, Alec and O'Brien, Kevin P.},
  journal = {Sci. Adv.},
  volume = {10},
  pages = {eado9094},
  year = {2024},
  doi = {10.1126/sciadv.ado9094}
}

@article{Lu2023_2,
  title = {High-fidelity parametric beamsplitting with a parity-protected converter},
  author = {Lu, Yao and Maiti, Aniket and Garmon, John W. O. and Ganjam, Suhas and Zhang, Yaxing and Claes, Jahan and Frunzio, Luigi and Girvin, S. M. and Schoelkopf, Robert J.},
  journal = {Nat. Commun.},
  volume = {14},
  pages = {5767},
  year = {2023},
  doi = {10.1038/s41467-023-41104-0}
}

\end{document}